\documentclass[pdflatex,sn-mathphys-num]{sn-jnl}

\usepackage{graphicx}
\usepackage{multirow}
\usepackage{amsmath,amssymb,amsfonts}
\usepackage{amsthm}
\usepackage[title]{appendix}
\usepackage{xcolor}
\usepackage{textcomp}
\usepackage{booktabs}
\usepackage{algorithm}
\usepackage{algorithmicx}
\usepackage{algpseudocode}

\theoremstyle{thmstyleone}
\newtheorem{theorem}{Theorem}
\newtheorem{proposition}[theorem]{Proposition}
\newtheorem{lemma}[theorem]{Lemma}
\newtheorem{corollary}[theorem]{Corollary}
\newtheorem{conjecture}[theorem]{Conjecture}
\theoremstyle{thmstyletwo}

\theoremstyle{thmstylethree}

\newcommand{\F}{\mathbb{F}}
\newcommand{\Ftwo}{\mathbb{F}_2}
\newcommand{\Ftn}{\mathbb{F}_{2^n}}
\newcommand{\Fn}{\mathbb{F}_2^n}
\newcommand{\Tr}{\mathrm{Tr}}
\newcommand{\W}{\mathcal{W}}
\newcommand{\Z}{\mathcal{Z}}
\newcommand{\NB}{\mathcal{N}}
\newcommand{\cost}{\mathrm{cost}}
\newcommand{\aff}{\mathrm{A}}
\makeatletter\gdef\orcid#1{(ORCID: \href{https://orcid.org/#1}{#1})}\makeatother

\hypersetup{bookmarksdepth=2}

\begin{document}

\title[Testing CCZ-equivalence to permutations]{Testing CCZ-equivalence to permutations via
projections of Walsh zero spaces, with applications to APN functions}

\author*[1]{\fnm{Emanuele} \sur{Bellini}\,\orcid{0000-0002-2349-0247}}\email{emanuele.bellini@tii.ae}
\affil*[1]{\orgname{Technology Innovation Institute}, \orgaddress{\city{Abu Dhabi}, \country{United Arab Emirates}}}

\abstract{Whether almost perfect nonlinear (APN) permutations exist in even dimension $n\ge 8$ is a
long-standing open question, the big APN problem. The only example known in even dimension, on
six bits, was found from a quadratic APN function that is not itself a permutation, by
Carlet--Charpin--Zinoviev (CCZ) equivalence. It is therefore natural to test APN functions for CCZ-equivalence to a
permutation. This amounts to finding two complementary $n$-dimensional subspaces among the
zeros of the Walsh transform, which lie in $\F_2^{2n}$. A known necessary condition only uses
their projections to $\F_2^n$ and cannot decide every case. We give an exact test for functions
whose Walsh supports are affine, including all quadratic functions. It reconstructs the
subspaces from their projections, through searches for subspaces of $\F_2^n$ and for
constrained bilinear maps. On the benchmarked functions in
dimension twelve, it lists them $39$ to $55$ times faster than the generic search of
the commonly used sboxU library. We then study the functions $x^3+a^{-1}\mathrm{Tr}(a^3x^9)$,
which Göloğlu and Pavlů proved not to be CCZ-equivalent to permutations when $4$ divides $n$.
With their result, our computations show that this holds for every even $n\le 24$. In
dimension $18$ the necessary condition is satisfied, and the new test decides the case. We
conjecture this for every even dimension. We also compile $25$ classes of quadratic
APN functions in dimension ten from published instances, with equivalence certificates. No function in
these classes or in their switching closure is CCZ-equivalent to a permutation. All code and data are public.}



\maketitle

\section{Introduction}\label{sec:intro}

A function $F:\Fn\to\Fn$ is \emph{almost perfect nonlinear} (APN) if, for every nonzero
$a\in\Fn$ and every $b\in\Fn$, the equation $F(x+a)+F(x)=b$ has at most two solutions.
APN functions offer the best possible resistance to differential cryptanalysis
\cite{DBLP:conf/eurocrypt/Nyberg93}. They are also related to optimal objects in coding
theory, sequence design and finite geometry; see \cite{DBLP:books/cu/Carlet2021} for an
account. Many block ciphers, for instance substitution--permutation networks, need
invertible S-boxes. In odd dimension APN permutations are
plentiful: the Gold functions $x^{2^i+1}$ with $\gcd(i,n)=1$
\cite{DBLP:journals/tit/Gold68,DBLP:conf/eurocrypt/Nyberg93}, the Kasami \cite{DBLP:journals/iandc/Kasami71}, Welch
\cite{DBLP:journals/tit/Dobbertin99} and Niho \cite{DobbertinNiho99} functions, and the inverse function are APN permutations of $\Ftn$ for odd $n$. In even
dimension the situation is completely different. There is no APN permutation of $\F_2^4$
\cite{DBLP:journals/dam/Hou06}. No APN power function and no quadratic APN function is a
permutation in even dimension (see \cite{DBLP:books/cu/Carlet2021}). The only APN
permutation known in even dimension is the six-bit permutation found by Browning, Dillon,
McQuistan and Wolfe \cite{BrowningDMW10}. Whether APN permutations exist for some even
$n\ge 8$ is known as the \emph{big APN problem}.

Dillon's permutation was not found by a search among permutations. It was obtained as a
function CCZ-equivalent \cite{DBLP:journals/dcc/CarletCZ98} to the quadratic APN function
$x^3+x^{10}+ux^{24}$ over $\F_{2^6}$, which is not itself a permutation. Two lines of attack
on the big APN problem have therefore been pursued. The first constructs permutations with a
prescribed structure, such as the butterfly structure that underlies Dillon's permutation
\cite{DBLP:conf/crypto/PerrinUB16,DBLP:journals/tit/CanteautDP17,DBLP:journals/ccds/CanteautPT19}.
The second searches for APN functions that are not permutations but are CCZ-equivalent to
one. This second approach relies on a characterization due to Canteaut and Perrin
\cite{DBLP:journals/ffa/CanteautP19}. A function $F$ is CCZ-equivalent to a permutation if and only if
the set of zeros of its Walsh transform contains two $n$-dimensional subspaces with trivial
intersection.

\subsection{Related work}\label{sec:related}

\paragraph{Constructions of permutations.}
Perrin, Udovenko and Biryukov \cite{DBLP:conf/crypto/PerrinUB16} decomposed Dillon's
permutation as an open butterfly on $\F_{2^3}^2$. Canteaut, Duval and Perrin
\cite{DBLP:journals/tit/CanteautDP17} generalized the construction to all dimensions $4k+2$,
obtaining differential uniformity four for $k>1$. Canteaut, Perrin and Tian
\cite{DBLP:journals/ccds/CanteautPT19} proved that a generalized butterfly is APN only on six
bits. Beierle, Brinkmann and Leander \cite{DBLP:journals/tit/BeierleBL21} classified APN
permutations with nontrivial linear self-equivalences in small dimension.

\paragraph{CCZ-equivalence to permutations.}
Göloğlu and Langevin \cite{DBLP:journals/ffa/GologluL21} proved that Gold functions are not
CCZ-equivalent to permutations in any even dimension. They proved the same for Kasami
functions in every dimension divisible by four. These results rest on a necessary condition:
if $F$ is CCZ-equivalent to a permutation, its non-bent components contain two subspaces whose
sum is the whole space. This condition follows directly from an observation of Browning et
al.\ \cite{BrowningDMW10}. Göloğlu and Pavlů \cite{DBLP:journals/ccds/GologluP21} turned it
into an algorithm, a depth-first search for the largest subspace contained in the set of
non-bent components. With it they showed that no member of the families of APN functions known
at that time is CCZ-equivalent to a permutation of $\F_2^{n}$ for even $n\le 12$, apart from
the known case $n=6$. They also checked random members of these families up to $n=18$, and
introduced the number of subspaces of the set of non-bent components as an EA-invariant.
Moreover, they proved that $x^3+a^{-1}\Tr(a^3x^9)$ is not CCZ-equivalent to a permutation when
$4\mid n$. As reported in \cite{DBLP:journals/ccds/GologluP21}, Canteaut and Perrin
independently designed a similar algorithm. Budaghyan, Ivkovic and Kaleyski
\cite{DBLP:journals/ccds/BudaghyanIK23} extended the result for $4\mid n$ to all quadratic
3-to-1 APN functions. Bénéteau, Goluboff, Kölsch and Vaghasiya
\cite{DBLP:journals/tit/BeneteauGKV26} showed that a quadratic function that is CCZ-equivalent
to a permutation has at least $3(2^{n/2}-1)$ non-bent components.

When the non-bent components do contain subspaces spanning $\Fn$, the necessary condition is
satisfied and therefore inconclusive. One then has to search for Walsh zero spaces in
$\F_2^{2n}$. Chase and Lisoněk \cite{DBLP:journals/ccds/ChaseL22} constructed Walsh zero spaces
for Gold functions in odd dimension. Bonnetain, Perrin and Tian
\cite{DBLP:conf/asiacrypt/BonnetainPT19} described generic algorithms for finding vector
spaces of a given dimension in a set; the test of \cite{DBLP:journals/ffa/CanteautP19} is one
of them. These algorithms, together with a test for CCZ-equivalence to a permutation, are
implemented in the library sboxU of Perrin et al.\ \cite{sboxU}. Most of the works cited below
used sboxU to check their functions.

\paragraph{Quadratic APN functions and switching.}
Edel and Pott \cite{DBLP:journals/amco/EdelP09} introduced the switching construction. With it
they found, in dimension six, the only known APN function that is CCZ-inequivalent to both
power and quadratic functions.

In dimension eight the only known APN function that is not quadratic is the Kasami function
$x^{57}$. The known quadratic APN functions form at least $3{,}808{,}491$ CCZ-classes
\cite[Theorem~5.1]{DBLP:journals/corr/abs-2508-04644}. Each item below gives the number of
classes first found in the given source. The counts of the first four items are those of
\cite[Section~4]{DBLP:journals/tit/BeierleL22}.
\begin{itemize}
\item $22$: the quadratic functions in the list of Edel and Pott
\cite{DBLP:journals/amco/EdelP09}, which collects the earlier constructions and includes the
Gold functions $x^3$ and $x^9$.
\item $8{,}157$: Yu, Wang and Li \cite{DBLP:journals/dcc/YuWL14}, with a matrix method.
\item $10$: Weng, Tan and Gong \cite{WengTG13}.
\item $2$: the family of Taniguchi \cite{DBLP:journals/dcc/Taniguchi19}.
\item $12{,}921$: Beierle and Leander \cite{DBLP:journals/tit/BeierleL22}. They found $12{,}733$
classes by a recursive tree search among functions with prescribed linear self-equivalences,
and $188$ by switching.
\item $5{,}412$: Yu and Perrin \cite{DBLP:journals/ccds/YuP22}, with the same matrix method.
\item $6{,}368$: Beierle, Leander and Perrin \cite{DBLP:journals/dcc/BeierleLP22}, by extending
quadratic APN functions in dimension seven.
\item $3{,}775{,}599$: Beierle, Langevin, Leander, Polujan and Rasoolzadeh
\cite{DBLP:journals/corr/abs-2508-04644,BeierleLLPR25data}, by extending vectorial bent
functions.
\end{itemize}
The first seven items give the $32{,}892$ classes known before
\cite{DBLP:journals/corr/abs-2508-04644}. Beierle et al.\ estimate the total number of classes
at about six million. In a recent preprint, Kuznetsov \cite{Kuznetsov26} reports four further
classes, which are not included in the count above. In \cite{DBLP:journals/tit/BeierleL22}
and \cite{DBLP:journals/corr/abs-2508-04644}, none of the new functions is CCZ-equivalent to a
permutation. For the known infinite families, see \cite{DBLP:journals/ccds/GologluP21}. We do
not repeat these tests in dimension eight.

In dimension ten the known APN functions that are not quadratic are the Kasami function
$x^{57}$ and the Dobbertin function $x^{339}$ \cite{Dobbertin01}. Section~\ref{sec:exp-known}
shows that the quadratic APN instances we considered form exactly $25$ CCZ-classes. For some
families we took all parameters and for others random samples, and for
\cite{DBLP:journals/tit/ZhengKLPT22} we used the instances of the preprint version
(Section~\ref{sec:exp-known}). We use the names F3, F4,
F12 and F13 of the families in \cite[Table~3]{DBLP:journals/ccds/Kaleyski21}. Each item below
gives a set of classes and the sources whose functions lie in them.
\begin{itemize}
\item $2$ classes: the Gold functions $x^3$ and $x^9$. The class of $x^3$ also contains
members of the relative-trace family of Zheng, Kan, Li, Peng and Tang
\cite{DBLP:journals/tit/ZhengKLPT22}.
\item $2$ classes: the family F3 of Budaghyan and Carlet \cite{DBLP:journals/tit/BudaghyanC08}.
They also contain Carlet's bivariate family (see
\cite[Table~2.1]{DBLP:journals/combtheory/GologluK25}) and members of the family of
\cite{DBLP:journals/tit/ZhengKLPT22}.
\item $2$ classes: the family F4, $x^3+a^{-1}\Tr(a^3x^9)$, of Budaghyan, Carlet and Leander
\cite{DBLP:journals/ffa/BudaghyanCL09}.
\item $6$ classes: the family F12 of Taniguchi \cite{DBLP:journals/dcc/Taniguchi19}.
\item $5$ classes: the family F13 of Budaghyan, Helleseth and Kaleyski
\cite{DBLP:journals/tit/BudaghyanHK20}, with parameters in $\F_4$. They also contain the two
biprojective families of Göloğlu \cite{DBLP:journals/tit/Gologlu22} and the first family of
Calderini, Li and Villa \cite{DBLP:journals/ffa/CalderiniLV23}. Two of the five instances of
Beierle and Leander \cite{DBLP:journals/tit/BeierleL22} lie in them. So do two instances and
some members of the family of \cite{DBLP:journals/tit/ZhengKLPT22}, and the second instance of
Budaghyan, Ivkovic and Kaleyski \cite[Table~2]{DBLP:journals/ccds/BudaghyanIK23}.
\item $3$ classes: the other three instances of \cite{DBLP:journals/tit/BeierleL22}. One of them
also contains the third instance of \cite{DBLP:journals/tit/ZhengKLPT22}.
\item $5$ classes: the other five instances of \cite[Table~2]{DBLP:journals/ccds/BudaghyanIK23},
obtained by adding five to seven terms with coefficients in $\F_4$ to $x^3$, $x^9$ or
$x^{33}$. The class of the first one also contains the function of Li, Zhou, Li and Qu
\cite{DBLP:journals/tit/LiZLQ22}. It also contains the second family of
\cite{DBLP:journals/ffa/CalderiniLV23}.
\end{itemize}
Budaghyan, Ivkovic and Kaleyski checked with sboxU that none of their functions is
CCZ-equivalent to a permutation. Göloğlu and Kölsch \cite{DBLP:journals/combtheory/GologluK25}
determined the equivalences between the biprojective families. Yu, Li, Ichanska and Kaleyski
\cite{YuLIK25} classified the quadratic APN functions in dimension ten with coefficients in
$\F_2$. There are three classes, those of $x^3$, $x^9$ and $x^3+\Tr(x^9)$, the last one in F4.
Yu, Kaleyski, Budaghyan and Li \cite{DBLP:journals/ffa/YuKBL20} did the same for $n\le 9$. The
same classification is the case $(n,m)=(10,1)$ of the QAM search of Ichanska, Berg, Kaleyski and
Yu \cite{DBLP:journals/ccds/IchanskaBKY26}. Their search for $(n,m)=(10,2)$ is only partial,
and the new APN function it reports is in dimension eight, so it adds no class in dimension
ten.

Budaghyan, Calderini and Villa \cite{DBLP:journals/ffa/BudaghyanCV20} determined the
equivalences between the known families in small dimension. Kaleyski
\cite{DBLP:journals/ccds/Kaleyski21} surveyed the invariants for telling classes apart. Kalgin and Idrisova \cite{DBLP:journals/ccds/KalginI23}
classified the quadratic APN functions in seven variables. This classification is reproduced
in the data of Beierle et al.\ \cite{BeierleLLPR26data}, which list the EA-classes of quadratic
$(n,m)$-APN functions for small $n\le m$. Recently, Göloğlu and Kölsch \cite{GologluK26} gave a
large family of APN permutations in every odd dimension divisible by three. They also gave
quadratic APN functions in even dimension that are not bijective. Their functions are defined
on $\F_{2^m}^3$, so none of them has dimension ten.

\subsection{Our contribution}\label{sec:contribution}

We study the big APN problem computationally, with an emphasis on exact methods whose results
can be checked independently. Our main contributions are the projection test and its
application to the family $x^3+a^{-1}\Tr(a^3x^9)$ up to dimension $24$. The other
contributions concern dimensions eight to ten.
\begin{enumerate}
\item We consider functions whose Walsh supports are affine subspaces. These include all
quadratic functions. We parametrize the $n$-dimensional subspaces of their Walsh zeros by
three objects: a subspace $B$ of the components with Walsh zeros, a subspace $P$ of $\Fn$ of the same
dimension, and a bilinear map $B\times P\to\Ftwo$ (Theorem~\ref{thm:dual}). The resulting test
for CCZ-equivalence to a permutation, which we call the projection test, searches for $B$ and
$P$ in $\Fn$ instead of searching for subspaces of $\F_2^{2n}$. It complements the necessary
condition of \cite{DBLP:journals/ffa/GologluL21,DBLP:journals/ccds/GologluP21}
(Proposition~\ref{cor:counting}): it decides exactly the cases in which that condition is
satisfied and therefore inconclusive. An example is the Gold functions in dimension ten, whose
non-bent components contain eleven $5$-dimensional subspaces that pairwise span $\F_2^{10}$.
We compared it with the generic search of the sboxU library. Listing all Walsh zero spaces
is $5$ to $18$ times faster in dimension ten, and $39$ to $55$ times faster in dimension
twelve (Section~\ref{sec:exp-ccz}).
\item The family $x^3+a^{-1}\Tr(a^3x^9)$ in dimensions $n=2m$ with $m$ odd is not covered by
the theorem of \cite{DBLP:journals/ccds/GologluP21}. For these dimensions we show that the
non-bent components contain a line of the Desarguesian spread whenever $a$ is a cube
(Proposition~\ref{prop:Fa-line}). So their dimension alone cannot exclude a permutation. We
decide these cases for $n\in\{6,10,14,18,22\}$ by computation, and conclude that no member of
the family is CCZ-equivalent to a permutation for even $n\le 24$ (Corollary~\ref{cor:Fa-20}).
In dimension $18$ the necessary condition is satisfied, and the case is decided by
Theorem~\ref{thm:dual}. Dimension $22$ needs a subspace search that exploits the symmetries of
the family. We state a conjecture that would settle the family in all even dimensions
(Conjecture~\ref{conj:Fa}). The case $n=22$, computed after the conjecture was formed, agrees
with it.
\item We compile $25$ classes of quadratic APN functions in dimension ten from the published
instances we could examine, and test them. With an
explicit EA-equivalence test we show that the published instances we could obtain lie in
these $25$ classes. These are the instances of the preprint version of
\cite{DBLP:journals/tit/ZhengKLPT22}, those of the bivariate families of
\cite{DBLP:journals/tit/Gologlu22,DBLP:journals/tit/LiZLQ22,DBLP:journals/ffa/CalderiniLV23},
and the second instance of \cite[Table~2]{DBLP:journals/ccds/BudaghyanIK23} as printed there.
In particular, the function of \cite{DBLP:journals/tit/LiZLQ22} is EA-equivalent to the first
instance of \cite[Table~2]{DBLP:journals/ccds/BudaghyanIK23}. No class is CCZ-equivalent to a
permutation. For the families considered in \cite{DBLP:journals/ccds/GologluP21} this confirms
their result independently, and our values of the largest subspace dimension agree with
theirs. The classes found after that work were tested in \cite{DBLP:journals/tit/BeierleL22}
and \cite{DBLP:journals/ccds/BudaghyanIK23}, so our test is an independent confirmation for
them as well. Of the $25$ classes, $18$ are excluded by the dimension bound, $4$ by the
spanning condition, and $3$ need Theorem~\ref{thm:dual}.
\item We compute the quadratic switching closure of the source functions listed in
Table~\ref{tab:switching}, in dimensions six, eight, nine and ten, with a certified equivalence
for every neighbour, and we test every quadratic class found. In dimensions eight to ten the
resulting quadratic classes have no non-quadratic switching neighbours. In dimension ten the
$25$ classes are closed under switching. In
dimensions nine and ten this extends the computations of \cite{DBLP:journals/tit/BeierleL22},
which were performed in dimensions seven and eight.
\item We search for new quadratic APN functions in dimension ten by symmetry-restricted sampling
of univariate polynomials and by the tree search of \cite{DBLP:journals/tit/BeierleL22} with a
large randomized campaign. Every function found is tested individually, and none is
CCZ-equivalent to a permutation. A search over bilinear forms in dimensions six to eight shows
that unstructured search does not scale.
\item We explain why the inverse function is not APN in even dimension. The sets of four
points on which it is affine, and which therefore violate the APN condition, are exactly the
two-dimensional subspaces $a\F_4$, $a\neq 0$ (Proposition~\ref{prop:inverse}). We then show
for $n=8$ that swapping two entries of the lookup table of the inverse function $I$ never
brings it closer to the APN property, measured by the cost~\eqref{eq:cost}. Here a swap
exchanges the outputs $I(i)$ and $I(j)$ of two inputs $i\neq j$. We also enumerate three
structured families of permutations: butterflies whose keyed permutations are built from power
maps, cyclotomic maps and switched inverse functions. None of them contains a permutation of
$\F_2^8$ closer to the APN property than the inverse function. The butterflies of this kind
contain none of $\F_2^{10}$ either (Sections~\ref{sec:inverse} and~\ref{sec:constructions}).
\end{enumerate}
We find no APN permutation in dimension eight or ten. We believe that the tools, in particular
the projection test, are useful for future searches. We make all code and data available at
\url{https://github.com/peacker/ccz-projection-test-apn}, together with a script that compares
every number reported here with the stored results.

\subsection{Outline}\label{sec:outline}

Section~\ref{sec:prelim} recalls definitions.
Section~\ref{sec:inverse} studies the inverse function and local search around it, and
Section~\ref{sec:constructions} structured families of permutations.
Section~\ref{sec:switching} recalls exact switching and describes how we compute the
switching closure. Section~\ref{sec:ccz} contains the parametrization of Walsh zero spaces and
the resulting test for CCZ-equivalence to a permutation, and Section~\ref{sec:family}
applies it to the family $x^3+a^{-1}\Tr(a^3x^9)$. Section~\ref{sec:experiments}
reports the experiments, and Section~\ref{sec:conclusion} concludes with open questions.

\section{Preliminaries}\label{sec:prelim}

We identify $\Fn$ with the finite field $\Ftn$ whenever a field structure is needed. The
components of $F$ are then also written $x\mapsto\Tr(\beta F(x))$ with $\beta\in\Ftn$. The
field coordinate $\beta$ and the bit coordinate $b$ of the same component are related by the
dual basis of the trace form; the repository contains this conversion. In our
computations $\Ftn$ is represented in a polynomial basis modulo a fixed primitive polynomial,
listed in the accompanying repository. For $x,y\in\Fn$ we write $x\cdot y$ for the usual
scalar product, and $\Tr$ denotes the absolute trace of $\Ftn$. An \emph{$(n,n)$-function}
is a map $F:\Fn\to\Fn$; its \emph{components} are the Boolean functions
$F_b(x)=b\cdot F(x)$, $b\neq 0$.

\paragraph{Differential properties.}
For $a,b\in\Fn$ let $\delta_F(a,b)=\#\{x\in\Fn: F(x+a)+F(x)=b\}$. The values
$\delta_F(a,b)$ are even, and $F$ is APN if $\delta_F(a,b)\le 2$ for all $a\neq 0$ and all
$b$. To measure how far $F$ is from being APN we use
\begin{equation}\label{eq:cost}
\cost(F)=\sum_{a\neq 0}\sum_{b}\binom{\delta_F(a,b)/2}{2},
\end{equation}
which vanishes exactly when $F$ is APN. An entry $\delta_F(a,b)=4$ contributes $1$, and an
entry $6$ contributes $3$. A \emph{$2$-flat} is an affine subspace of dimension two,
$\{x,x+a,x+b,x+a+b\}$ with $a,b$ linearly independent. We say that $F$ is \emph{affine on} a
$2$-flat if $F(x)+F(x+a)+F(x+b)+F(x+a+b)=0$. We write $\aff(F)$ for the number of $2$-flats on
which $F$ is affine. A function is APN if and only if it is affine on no $2$-flat.

\paragraph{Walsh transform.}
The Walsh transform of $F$ is
$\W_F(a,b)=\sum_{x\in\Fn}(-1)^{a\cdot x+b\cdot F(x)}$ for $a,b\in\Fn$. The \emph{Walsh zeros}
of $F$ form the set $\Z_F=\{(a,b)\neq(0,0):\W_F(a,b)=0\}$, which always contains
$\{(a,0):a\neq 0\}$. For $b\neq0$ the \emph{Walsh support} of the component $F_b$ is
$S_b=\{a:\W_F(a,b)\neq 0\}$. The component is \emph{bent} if $|\W_F(a,b)|=2^{n/2}$ for all
$a$. We write $\NB_F=\{b\neq 0: S_b\neq\Fn\}$ for the set of components that have a Walsh zero.
A bent component has no Walsh zero. For plateaued components, in particular for quadratic
ones, the converse holds, so for quadratic $F$ the set $\NB_F$ is the set of non-bent
components. This is not true in general. For example, the Boolean function on $\F_2^3$ that is
$1$ at exactly one point has no Walsh zero but is not bent. All our proofs use only the
definition of $\NB_F$. In general statements we call its elements the components with Walsh
zeros. For quadratic functions, where they are exactly the non-bent components, we also use
that name. If $F$ is quadratic, the
component $F_b$ has a linear space
$L_b=\{y: F_b(x+y)+F_b(x)+F_b(y)+F_b(0)=0\ \text{for all}\ x\}$, and $S_b$ is a coset of
$L_b^\perp$, so that $|\W_F(a,b)|\in\{0,2^{(n+\dim L_b)/2}\}$; see
\cite{DBLP:books/cu/Carlet2021}.

\paragraph{Equivalences.}
Two functions $F,G$ are \emph{CCZ-equivalent} if an affine permutation of $\F_2^{2n}$ maps
the graph $\{(x,F(x))\}$ of $F$ onto the graph of $G$ \cite{DBLP:journals/dcc/CarletCZ98}. They
are \emph{EA-equivalent} if $G=A_1\circ F\circ A_2+A_3$ with $A_1,A_2$ affine permutations and
$A_3$ affine. CCZ-equivalence preserves the
differential spectrum and the extended Walsh spectrum, hence the APN property. For quadratic
APN functions CCZ-equivalence coincides with EA-equivalence \cite{DBLP:journals/jalc/Yoshiara12}.
The following characterization is the basis of our tests.

\begin{proposition}[\cite{DBLP:journals/ffa/CanteautP19}]\label{prop:cp}
$F$ is CCZ-equivalent to a permutation if and only if $\Z_F\cup\{0\}$ contains two
$n$-dimensional linear subspaces $U,V$ with $U\cap V=\{0\}$.
\end{proposition}

Given such $U=\langle u_1,\dots,u_n\rangle$ and $V=\langle v_1,\dots,v_n\rangle$, a
permutation in the CCZ-class of $F$ is $G=G_2\circ G_1^{-1}$ with
$G_1(x)=(u_i\cdot(x,F(x)))_{i}$ and $G_2(x)=(v_j\cdot(x,F(x)))_{j}$. Indeed, the linear map with
rows $u_1,\dots,u_n,v_1,\dots,v_n$ maps the graph of $F$ onto a set whose Walsh transform
vanishes on both coordinate axes, so that both projections of this set are bijective. Following
\cite{DBLP:journals/ccds/ChaseL22} we call an $n$-dimensional subspace of $\Z_F\cup\{0\}$ a
\emph{Walsh zero space} of $F$.

\paragraph{Invariants.}
To tell CCZ-classes of quadratic APN functions apart we use the ortho-derivative. For
$a\neq 0$, the image of $x\mapsto F(x+a)+F(x)+F(a)+F(0)$ is a hyperplane, and $\pi_F(a)$ is
the nonzero vector orthogonal to it. The differential spectrum and the extended Walsh
spectrum of $\pi_F$ are EA-invariants of $F$
\cite{DBLP:journals/tit/CanteautCP22,DBLP:journals/ccds/Kaleyski21}, hence CCZ-invariants for
quadratic APN functions \cite{DBLP:journals/jalc/Yoshiara12}.
These invariants are not complete. Two functions with different invariants are
inequivalent, but equal invariants do not prove equivalence. We say that a function lies in a
class only when we have proved the equivalence by an explicit certificate
(Section~\ref{sec:exp-known}). Otherwise we only say that it has the invariant of that class.
All CCZ-equivalence tests are applied to every function individually
(Section~\ref{sec:experiments}). Our conclusions therefore do not depend on the completeness of
the invariant.

\section{The inverse function and local search}\label{sec:inverse}

The cost \eqref{eq:cost} has a simple geometric meaning.

\begin{lemma}\label{lem:cost}
For every $(n,n)$-function $F$, $\cost(F)=3\,\aff(F)$.
\end{lemma}

\begin{proof}
Fix $a\neq 0$ and $b$. The solutions of $F(x+a)+F(x)=b$ come in $\delta_F(a,b)/2$ pairs
$\{x,x+a\}$, and every unordered pair of distinct such pairs $\{x,x+a\}$, $\{y,y+a\}$
determines the $2$-flat $\{x,x+a,y,y+a\}$, on which $F$ is affine. Conversely, let $F$ be
affine on a $2$-flat $E$ with direction space $\{0,a,a',a+a'\}$. For each of the three
nonzero directions $c$ of $E$, the two pairs of $E$ in direction $c$ have the same derivative
value. Hence every $2$-flat on which $F$ is affine is counted exactly three times in
\eqref{eq:cost}, once for each nonzero direction.
\end{proof}

Let $I(x)=x^{2^n-2}$ be the inverse function, with $I(0)=0$. For even $n$ the field $\Ftn$
contains $\F_4=\{0,1,\omega,\omega^2\}$, and $\Ftn$ is a vector space over $\F_4$; we call the
sets $a\F_4$, $a\neq 0$, the \emph{$\F_4$-lines through the origin}. There are
$(2^n-1)/3$ of them. The following proposition is a geometric restatement of Nyberg's
computation of the differential spectrum of $I$ \cite{DBLP:conf/eurocrypt/Nyberg93}.

\begin{proposition}\label{prop:inverse}
Let $n$ be even. The $2$-flats on which $I$ is affine are exactly the $\F_4$-lines through
the origin. In particular $\aff(I)=(2^n-1)/3$ and $\cost(I)=2^n-1$.
\end{proposition}

\begin{proof}
For $a\neq 0$ the equation $I(x)+I(x+a)=b$ has at most four solutions, and four solutions
only for $b=a^{-1}$ \cite{DBLP:conf/eurocrypt/Nyberg93}. The four solutions are
$x\in a\F_4$: this is clear for $x\in\{0,a\}$, and for $x=a\omega$ we get
$(a\omega)^{-1}+(a\omega^2)^{-1}=a^{-1}(\omega^2+\omega)=a^{-1}$, using $\omega+1=\omega^2$,
and symmetrically for $x=a\omega^2$. So the $2$-flats on which $I$ is affine are the sets
$a\F_4$, each of which is a $2$-dimensional linear subspace with nonzero directions
$a,a\omega,a\omega^2$; by Lemma~\ref{lem:cost} there are $\cost(I)/3=(2^n-1)/3$ of them.
\end{proof}

Proposition~\ref{prop:inverse} says that the inverse function is close to APN in a very
regular way: it fails the APN condition only on the subspaces $a\F_4$, and all of them
contain the point $0$. A modification of $I$ that produces an APN permutation must
destroy all these flats simultaneously without creating new ones. Changing the value $I(0)=0$ to
any other value already destroys all of them, since each of these flats contains $0$. The
result is no longer a permutation, so other values must change as well, and the question is
how many new flats are created.

\paragraph{Local search.}
For $n=8$ we measured how $\cost$ changes when two entries of the lookup table are swapped,
i.e.\ under the moves $S\mapsto S\circ(i\;j)$, which exchange the outputs $S(i)$ and $S(j)$.
With an incremental update of the difference table, one move is evaluated in $O(2^n)$
operations. All $\binom{256}{2}=32{,}640$ transpositions of $I$ can then be scanned in a few
milliseconds. The result is that no swap of two entries improves $I$. No transposition
decreases the cost. The $255$ transpositions that leave it unchanged are exactly those
involving the position $0$. Every other transposition increases the cost by at least $222$.
None of the $255$ permutations reached by the cost-preserving transpositions can be improved by
a further transposition. We also ran simulated annealing started at $I$, with plain
transpositions or with moves that preserve the involution property of $I$; recall that
Dillon's permutation is CCZ-equivalent to an involution. It never found a permutation of cost
below $255$ (Section~\ref{sec:exp-inverse}). For comparison, simulated annealing started from
random permutations of $\F_2^8$ does not get below costs between $4{,}776$ and $4{,}926$.

\section{Structured families of permutations}\label{sec:constructions}

We considered three families of permutations that contain, or are related to, the
known APN permutations. We searched them exhaustively in dimension eight, and the first family
also in dimension ten. All results are negative and are collected in
Table~\ref{tab:constructions}.

\paragraph{Butterflies with power-based keyed permutations.}
Let $n=2k$ and let $R_z$, $z\in\F_{2^k}$, be a family of permutations of $\F_{2^k}$. The
\emph{open butterfly} \cite{DBLP:conf/crypto/PerrinUB16}
\[
H(x,y)=\bigl(R_{t}(y),\,t\bigr),\qquad t=R_y^{-1}(x),
\]
is a permutation of $\F_{2^k}^2$ for every such family. Dillon's permutation is an open
butterfly with $k=3$ and $R_z(x)=(x+\alpha z)^3+z^3$ \cite{DBLP:conf/crypto/PerrinUB16}. The
generalized butterflies of \cite{DBLP:journals/tit/CanteautDP17,DBLP:journals/ccds/CanteautPT19}
use $R_z(x)=(x+\alpha z)^e+z^e$ with $e=2^i+1$, which requires $k$ odd. For $n=8$ the exponent $3$ is not
available, since $x^3$ is not a permutation of $\F_{16}$. We therefore allowed
\[
R_z(x)=(x+\alpha z)^e+\beta z^f.
\]
Here $e$ runs over one representative of each cyclotomic class of exponents coprime to
$2^k-1$. Further, $f$ runs over $0,\dots,2^k-2$ (with $z^0$ read as $1$ for $z\neq0$ and as $0$ for
$z=0$), and $\alpha,\beta$ run over all of $\F_{2^k}$, including $0$. Every such tuple gives a
permutation, and we count tuples, not distinct lookup tables. For $k=3$ six tuples give APN
permutations; we did not check how many of them are equivalent. For $k=4$ and $k=5$ the smallest costs are $8{,}280$ and $71{,}424$, compared
with $255$ and $1{,}023$ for the inverse function.

\paragraph{Cyclotomic maps.}
For $\ell\mid 2^n-1$ and $s=(2^n-1)/\ell$, a cyclotomic map is $S(x)=x^r\,h(x^s)$ with $h$
defined on the $\ell$-th roots of unity. In other words, it is a power function multiplied by
a constant that depends on the coset of $x$ in $\Ftn^*$ modulo the subgroup of index $\ell$.
Both $x^3$, which is APN but not a permutation for even $n$, and $I$ belong to such families.
For $n=8$ and $\ell=3$, each $\F_4$-line $a\F_4^*$ of Proposition~\ref{prop:inverse} meets
each of the three cosets in exactly one point, because $\omega=\alpha^{85}$ and $85$ is not
divisible by three. For $n=8$ and $\ell=3$ we took one representative $r$ of each cyclotomic
class with $\gcd(r,85)=1$, $24$ classes in all, and all $h$ with $h(1)=1$. We kept the maps
that are permutations. None of them has cost below $255$, and the nontrivial choices of $h$
applied to the inverse function have cost at least $4{,}080$.

\paragraph{Switched inverse functions.}
Finally we considered $S(x)=I(x)+\gamma\Tr(\delta x^d)$ for all exponents $d$, up to
cyclotomic equivalence, and all $\gamma,\delta\in\F_{2^8}^*$. Such a switching modifies $I$ on
a large set at once. Among the $33{,}660$ parameter tuples that give a permutation, $32{,}385$
give cost $255$, the cost of $I$. Every other one has cost at least $4{,}155$.

\begin{table}[t]
\caption{Structured families of permutations (Section~\ref{sec:constructions}). Butterflies use
$R_z(x)=(x+\alpha z)^e+\beta z^f$, switched inverses $I+\gamma\Tr(\delta x^d)$. ``Exp.\
classes'' is the number of cyclotomic classes of $e$, $r$ or $d$; ``Perms.'' counts
parameter tuples that give a permutation, without merging equal or equivalent tables; for the
switched inverse the second-smallest cost is also given. The
cost of $I$ is $2^n-1$; a cost of $0$ means APN.}\label{tab:constructions}
\begin{tabular*}{\textwidth}{@{\extracolsep\fill}llrrrr}
\toprule
Family & $n$ & Exp.\ classes & Perms. & APN & Min.\ cost \\
\midrule
Butterflies & 6 & 2 & 896 & 6 & 0 \\
 & 8 & 2 & 7,680 & 0 & 8,280 \\
 & 10 & 6 & 190,464 & 0 & 71,424 \\
Cyclotomic $x^rh(x^{85})$ & 8 & 24 & 346,800 & 0 & 255 \\
Switched inverse & 8 & 35 & 33,660 & 0 & 255; 4,155 \\
\botrule
\end{tabular*}
\end{table}

These experiments indicate that, in contrast with dimension six, the natural structured
generalizations of the known constructions do not come close to the APN property in
dimensions eight and ten. The inverse function has the smallest cost we observed for a
permutation. The modifications of it that we tried either keep its cost or have a much
larger one. This motivates the approach of the following
sections, which searches for APN functions that are not permutations and tests them for
CCZ-equivalence to a permutation.

\section{Exact switching and switching neighbourhoods}\label{sec:switching}

Two functions $F,G$ are \emph{switching neighbours} if $G=F+uf$ for a Boolean function $f$
and a nonzero $u\in\Fn$ \cite{DBLP:journals/amco/EdelP09}. If $F$ is APN, all its APN
switching neighbours can be found by linear algebra.

\begin{proposition}[\cite{DBLP:journals/amco/EdelP09}]\label{prop:switching}
Let $F$ be APN and $u\neq 0$. Then $F+uf$ is APN if and only if
\[
f(x)+f(x+a)+f(y)+f(y+a)=0
\]
for all $a\neq 0$ and all $x,y$ such that $F(x)+F(x+a)+F(y)+F(y+a)=u$.
\end{proposition}

\begin{proof}[Sketch]
$F+uf$ fails to be APN iff there are $a\neq0$ and two distinct pairs $\{x,x+a\}$,
$\{y,y+a\}$ with equal derivatives, i.e.\
$D_aF(x)+D_aF(y)=u\,(D_af(x)+D_af(y))$. Since $F$ is APN the left-hand side is nonzero, so
this happens iff $D_aF(x)+D_aF(y)=u$ and $D_af(x)+D_af(y)=1$.
\end{proof}

The conditions are linear in the truth table of $f$. So the set $K(F,u)$ of all $f$ with
$F+uf$ APN is a subspace of the space of Boolean functions. We compute it exactly by
Gaussian elimination over $\Ftwo$. In dimension ten the system has $2^{10}$ unknowns and
up to a few hundred thousand equations; with rows packed into machine words it is solved
in about a quarter of a second.

$K(F,u)$ always contains the affine Boolean functions, and it contains the components $f=b\cdot
F$ with $b\cdot u=0$, for which $F+uf=L\circ F$ with the invertible linear map
$L(y)=y+u\,(b\cdot y)$. We call the span $T(F,u)$ of these functions the \emph{trivial} part
of $K(F,u)$. If $f\in K(F,u)$ and $t\in T(F,u)$, then $F+u(f+t)$ is EA-equivalent to $F+uf$:
for $G=F+uf$ and $b\cdot u=0$ we have $b\cdot F=b\cdot G$, so $G+u(b\cdot F)=L\circ G$. Hence
the switching neighbours of $F$ in direction $u$ are, up to EA-equivalence, given by the
$2^q-1$ nonzero classes of the quotient $K(F,u)/T(F,u)$, where $q=\dim K(F,u)-\dim T(F,u)$.

\begin{lemma}\label{lem:nonquadratic}
Let $F$ be quadratic and $f\in K(F,u)$. Then $F+uf$ has algebraic degree at least three if
and only if the algebraic normal form of $f$ contains a monomial of degree at least three.
Consequently, let $K_{\le2}$ be the subspace of the elements of $K(F,u)$ of degree at most
two, and let $r=\dim K(F,u)-\dim K_{\le2}$. Then $r$ is the rank of the degree-$\ge3$ parts of
the algebraic normal forms of a basis of $K(F,u)$, and $F$ has $2^{\dim K(F,u)}-2^{\dim
K_{\le2}}$ switching neighbours of degree at least three in direction $u$. In particular such
neighbours exist if and only if $r>0$.
\end{lemma}

\begin{proof}
The coordinate functions of $F+uf$ are $F_i+u_if$; those of $F$ have degree at most two,
and $u$ has a nonzero coordinate. The map that sends $f$ to the degree-$\ge3$ part of its
algebraic normal form is linear on $K(F,u)$, with kernel $K_{\le2}$ and image of dimension
$r$.
\end{proof}

\paragraph{Switching closure.}
Starting from a list of APN functions, we compute $K(F,u)$ for every source $F$ and every
$u\neq 0$. For a power function $x^d$ it suffices to take $u$ modulo the $\gcd(d,2^n-1)$-th
powers, because of the equivalence $x\mapsto cx$. For each $u$ we record whether
non-quadratic solutions exist (Lemma~\ref{lem:nonquadratic}). We then form one function
$G=F+uf$ for each nonzero class of $K(F,u)/T(F,u)$. In our computations $q\le 5$ in dimension six
and $q\le 2$ in dimensions eight to ten, so this is cheap. Each quadratic $G$ is compared with the stored functions that have the same
ortho-derivative spectra, with the EA-equivalence test of Section~\ref{sec:exp-known}. This
test is exhaustive and returns an explicit certificate, which is checked separately. If $G$ is
equivalent to a stored function, it is discarded. Otherwise it becomes a new class, is
stored, and is in turn switched. EA-equivalent functions have EA-equivalent switching
neighbours: if $G=L_1\circ F\circ L_2+A$, then $L_1\circ(F+uf)\circ L_2+A=G+L_1(u)\,(f\circ L_2)$.
So it suffices to switch one function per class. The procedure ends when every
stored class has been switched. Non-quadratic neighbours are recorded but not switched further. The result is therefore the
exact set of quadratic APN classes reachable from the sources by switching paths whose
intermediate functions remain quadratic, and the classes in it are pairwise inequivalent. In
dimensions eight to ten no quadratic class in the closure has a non-quadratic neighbour, so
there the closure is also closed under unrestricted switching. In dimension six it is not, but
the known classification shows that all thirteen quadratic classes are reached. Every new class is tested for
CCZ-equivalence to a permutation with the test of Section~\ref{sec:ccz}.

In dimension six we started this procedure from three sources: $x^3$, the functions
$x^3+a^{-1}\Tr(a^3x^9)$ with $a\in\F_4^*$, and $x^3+\alpha(x^{17}+x^{18}+x^{20}+x^{24})$, where
$\alpha$ is a primitive element. It reaches all thirteen classes of quadratic APN functions.
Five of them have switching neighbours of degree three. Among them is the last source, whose
switching gives the non-quadratic APN function of Edel and Pott
\cite{DBLP:journals/amco/EdelP09}. Exactly one of the thirteen classes, the class of the Kim
function, is found to be CCZ-equivalent to a permutation by the test of Section~\ref{sec:ccz}.
The procedure thus rediscovers both known exceptional phenomena of dimension six, which serves
as a check of the implementation. The results in dimensions eight, nine and ten are reported
in Section~\ref{sec:exp-switching}.

\section{The projection test for CCZ-equivalence to a permutation}\label{sec:ccz}

By Proposition~\ref{prop:cp}, deciding whether $F$ is CCZ-equivalent to a permutation
amounts to finding all Walsh zero spaces of $F$, i.e.\ all $n$-dimensional subspaces of
$\Z_F\cup\{0\}\subseteq\F_2^{2n}$, and testing them for a pair with trivial intersection.
For a quadratic APN function in even dimension, $\Z_F$ contains about a quarter of
$\F_2^{2n}$. In dimension ten this is $262{,}911$ vectors for $x^3$. A generic search
\cite{DBLP:conf/asiacrypt/BonnetainPT19} has to find all $10$-dimensional subspaces of this set. In this
section we assume that the Walsh supports are affine. We show that a Walsh zero space is then
described by three objects: a subspace $B$ of $\NB_F\cup\{0\}$, a subspace $P$ of $\Fn$ of the same
dimension $m$, and a bilinear map $\Psi$ with $m^2$ binary entries. The searches for $B$ and $P$
take place in $\Fn$. We do not claim an improvement of the worst-case complexity of the
generic search; the gain is practical, and we measure it in Section~\ref{sec:exp-ccz}. Throughout, $\pi:\F_2^{2n}\to\Fn$, $\pi(a,b)=b$, denotes the
projection onto the second coordinate, and for a Walsh zero space $U$ we write
$U_0=\{a:(a,0)\in U\}$.

\subsection{Projections onto the components with Walsh zeros}

\begin{lemma}\label{lem:projection}
Let $U$ be a Walsh zero space of an $(n,n)$-function $F$. Then
$\pi(U)\subseteq\NB_F\cup\{0\}$, and $\dim U_0=n-\dim\pi(U)$.
\end{lemma}

\begin{proof}
If $b\neq0$ is not in $\NB_F$, then $S_b=\Fn$, so $\W_F(a,b)\neq 0$ for every $a$ and no
vector $(a,b)$ can lie in $U$. The second statement is the rank--nullity theorem
for $\pi|_U$, whose kernel is $U_0\times\{0\}$.
\end{proof}

\begin{proposition}[\cite{DBLP:journals/ffa/GologluL21,DBLP:journals/ccds/GologluP21}]\label{cor:counting}
If $F$ is CCZ-equivalent to a permutation, then $\NB_F\cup\{0\}$ contains two linear
subspaces $B_1,B_2$ with $B_1+B_2=\Fn$. In particular, if $M$ denotes the largest dimension
of a linear subspace contained in $\NB_F\cup\{0\}$, then $M\ge n/2$, and every Walsh zero
space $U$ that belongs to a pair with trivial intersection satisfies
$n-M\le\dim\pi(U)\le M$.
\end{proposition}

\begin{proof}
We recall the short argument for completeness. If $U\cap V=\{0\}$ for two Walsh zero spaces, then $U+V=\F_2^{2n}$, hence
$\pi(U)+\pi(V)=\Fn$, and both projections lie in $\NB_F\cup\{0\}$ by
Lemma~\ref{lem:projection}. Then $\dim\pi(U)+\dim\pi(V)\ge n$ with both dimensions at most
$M$.
\end{proof}

Proposition~\ref{cor:counting} is Condition~2 and Corollary~1 of
\cite{DBLP:journals/ccds/GologluP21}. The condition was used in
\cite{DBLP:journals/ffa/GologluL21} for Gold and Kasami functions. The number of subspaces of
$\NB_F\cup\{0\}$ is an EA-invariant
\cite{DBLP:journals/ccds/GologluP21,DBLP:journals/ccds/Kaleyski21}. For quadratic $F$,
Bénéteau, Goluboff, Kölsch and Vaghasiya \cite[Corollary~V.6]{DBLP:journals/tit/BeneteauGKV26}
show two further necessary conditions: $|\NB_F|\ge 3(2^{n/2}-1)$, and every
$(n/2)$-dimensional subspace meets $\NB_F$ in at least three points. In dimension ten the
quadratic APN functions considered below have $341$ non-bent components, so the bound on
$|\NB_F|$ does not exclude them. Budaghyan, Ivkovic and Kaleyski
\cite[Theorem~7]{DBLP:journals/ccds/BudaghyanIK23} prove $M\le n/2-1$ for every quadratic
3-to-1 APN function when $4\mid n$. Hence no such function is CCZ-equivalent to a
permutation. By \cite[Proposition~8]{DBLP:journals/ccds/BudaghyanIK23}, this applies for
$4\mid n$ to all APN monomials and to several infinite families, among them
$x^3+a^{-1}\Tr(a^3x^9)$. The case $n\equiv 2\pmod 4$, which contains dimension ten, is left
open there.

Computing $M$ settles every function with $M<n/2$. The computation is inconclusive when
$\NB_F\cup\{0\}$ contains subspaces spanning $\Fn$. In that case one has to decide whether
these subspaces are projections of Walsh zero spaces that intersect trivially. The rest of
this section shows how to do so.

\subsection{Walsh zero spaces of functions with affine supports}

We now assume that for every $b\in\NB_F$ the Walsh support is an affine subspace
$S_b=s_b+D_b$, with $D_b$ a linear subspace. We set $L_b=D_b^\perp$. This holds for every
quadratic function, with $L_b$ the linear space of the component $F_b$
(Section~\ref{sec:prelim}). Note that $y\cdot s_b$ does not depend on the choice of
$s_b\in S_b$ when $y\in L_b$.

\begin{theorem}\label{thm:dual}
Let $F$ have affine Walsh supports. Let $B\subseteq\NB_F\cup\{0\}$ and $P\subseteq\Fn$ be
linear subspaces of the same dimension $m$, and let $\Psi:B\times P\to\Ftwo$ be a bilinear
map. Let $\varphi:B\to\Fn$ be a linear map with $y\cdot\varphi(b)=\Psi(b,y)$ for all $b\in B$
and $y\in P$, and set
\[
U(B,P,\Psi)=\{(a,b): b\in B,\ a\in\varphi(b)+P^\perp\}.
\]
Then:
\begin{enumerate}
\item $U(B,P,\Psi)$ is an $n$-dimensional subspace of $\F_2^{2n}$ that does not depend on the
choice of $\varphi$, with $\pi(U)=B$ and $U_0=P^\perp$;
\item every $n$-dimensional subspace $U$ with $\pi(U)\subseteq\NB_F\cup\{0\}$ equals
$U(B,P,\Psi)$ for exactly one triple, namely $B=\pi(U)$, $P=U_0^\perp$ and
$\Psi(b,y)=y\cdot a$ for any $(a,b)\in U$;
\item $U(B,P,\Psi)$ is a Walsh zero space of $F$ if and only if for every $b\in
B\setminus\{0\}$ there exists $y\in P\cap L_b$ with $\Psi(b,y)\neq y\cdot s_b$.
\end{enumerate}
\end{theorem}

\begin{proof}
(1) A map $\varphi$ exists because $P^\perp$ has codimension $m$ and $\Fn/P^\perp$ is
dual to $P$. Two choices of $\varphi$ differ by a linear map $B\to P^\perp$, which does not
change the set $U(B,P,\Psi)$. The set is the image of the injective linear map
$(b,c)\mapsto(\varphi(b)+c,b)$ on $B\times P^\perp$, so it has dimension $m+(n-m)=n$. Its
projection is $B$ and its kernel is $P^\perp$.

(2) Let $U$ be given and $B=\pi(U)$, $m=\dim B$. By the rank--nullity theorem for $\pi|_U$,
whose kernel is $U_0\times\{0\}$, we have $\dim U_0=n-m$, so $P=U_0^\perp$ has dimension $m$. For $b\in B$ the fibre
$\{a:(a,b)\in U\}$ is a coset of $U_0=P^\perp$. So $y\cdot a$ is the same for all $a$ in the
fibre when $y\in P$, and $\Psi$ is well defined and bilinear. The fibres of $U$ and of
$U(B,P,\Psi)$ over each $b$ are the same coset of $P^\perp$, hence the two sets coincide.
Conversely $B$, $P$ and $\Psi$ are determined by $U$, which proves uniqueness.

(3) The fibre over $b=0$ is $P^\perp$, and $(a,0)\in\Z_F$ for all $a\neq 0$. For
$b\in B\setminus\{0\}$ the fibre $\varphi(b)+P^\perp$ consists of Walsh zeros if and only if
it is disjoint from $S_b=s_b+D_b$. Two cosets $x+X$ and $z+Y$ are disjoint if and only if
$x+z\notin X+Y$, and $P^\perp+D_b=(P\cap L_b)^\perp$. Hence the fibre consists of Walsh
zeros if and only if there exists $y\in P\cap L_b$ with $y\cdot(\varphi(b)+s_b)=1$, that is,
$\Psi(b,y)\neq y\cdot s_b$.
\end{proof}

In particular a Walsh zero space with projection $B$ requires a subspace $P$ of dimension
$\dim B$ that meets $L_b$ nontrivially for every $b\in B\setminus\{0\}$. For quadratic APN
functions in even dimension the non-bent components typically have $\dim L_b=2$, so this is a
strong condition.

\paragraph{Relation to previous work.}
Parts (1) and (2) of Theorem~\ref{thm:dual} are linear algebra: a subspace of
$\F_2^{2n}$ is determined by its projection, its kernel and a linear map between them. The new
ingredient is part (3). For affine supports the condition that a fibre consists of Walsh zeros
involves only $P\cap L_b$ and $s_b$. So the possible subspaces $P$ are cut out by the spaces
$L_b$, and $\Psi$ by equations that are linear in the typical case. The characterization of
Canteaut and Perrin \cite{DBLP:journals/ffa/CanteautP19} and the generic search of Bonnetain,
Perrin and Tian \cite{DBLP:conf/asiacrypt/BonnetainPT19} work with the set $\Z_F$ in
$\F_2^{2n}$. Chase and Lisoněk \cite{DBLP:journals/ccds/ChaseL22} construct Walsh zero spaces
of Gold functions explicitly. The necessary condition of
\cite{DBLP:journals/ffa/GologluL21,DBLP:journals/ccds/GologluP21} uses only the projections.
Our test starts from these projections and decides exactly which of them lift to Walsh zero
spaces.

\subsection{The algorithm}

Theorem~\ref{thm:dual} and Proposition~\ref{cor:counting} give Algorithm~\ref{alg:dual}, which we
call the \emph{projection test}, since it reconstructs the Walsh zero spaces from their
projections. All searches take place in $\Fn$ or in $B\times P$, never in $\F_2^{2n}$.

\begin{proposition}\label{prop:alg-correct}
Let $F$ have affine Walsh supports. Algorithm~\ref{alg:dual} returns a permutation if and
only if $F$ is CCZ-equivalent to a permutation, and every permutation it returns is
CCZ-equivalent to $F$.
\end{proposition}

\begin{proof}
If $U,V$ are Walsh zero spaces with $U\cap V=\{0\}$, then $M\ge n/2$ and
$n-M\le\dim\pi(U),\dim\pi(V)\le M$ by Proposition~\ref{cor:counting}. By
Theorem~\ref{thm:dual}(2) and (3), $U$ equals $U(\pi(U),P,\Psi)$ for a subspace $P$ and a map
$\Psi$ that the loops enumerate, and likewise for $V$. So both lie in $\mathcal{U}$, and the
algorithm returns a permutation. Conversely, every element of $\mathcal{U}$ is a Walsh zero
space by Theorem~\ref{thm:dual}(3). A pair with trivial intersection gives a permutation
CCZ-equivalent to $F$ by Proposition~\ref{prop:cp}.
\end{proof}

\begin{algorithm}[t]
\caption{Projection test: CCZ-equivalence to a permutation for functions with affine Walsh supports}\label{alg:dual}
\begin{algorithmic}[1]
\State for each $b\neq 0$ with $S_b\neq\Fn$ compute $s_b$ and a basis of $L_b$ (from $\W_F$, or
from the bilinear form if $F$ is quadratic)
\State $M\gets$ largest dimension of a subspace of $\NB_F\cup\{0\}$
\If{$M<n/2$} \Return \emph{not CCZ-equivalent to a permutation} \EndIf
\State $\mathcal{U}\gets\emptyset$
\ForAll{subspaces $B\subseteq\NB_F\cup\{0\}$ with $n-M\le\dim B\le M$}
  \ForAll{subspaces $P$, $\dim P=\dim B$, with $P\cap L_b\neq\{0\}$ for all $b\in B\setminus\{0\}$}
    \State enumerate the bilinear maps $\Psi$ satisfying condition (3) of Theorem~\ref{thm:dual}
    \State add every $U(B,P,\Psi)$ to $\mathcal{U}$
  \EndFor
\EndFor
\If{$U,V\in\mathcal{U}$ with $U\cap V=\{0\}$ exist} \Return the permutation $G_2\circ G_1^{-1}$ of
Section~\ref{sec:prelim}
\Else\ \Return \emph{not CCZ-equivalent to a permutation} \EndIf
\end{algorithmic}
\end{algorithm}

The subspaces $B$ are found with a standard vector space search in the set
$\NB_F\cup\{0\}\subseteq\Fn$, where every subspace is generated once through its reduced
echelon basis. The subspaces $P$ are built by a depth-first search. At each step it takes the
first $b$ whose $L_b$ is not yet met by the current partial basis, and adds one of the nonzero
vectors of $L_b$. Once all $L_b$ are met, it completes the basis in all possible ways. Every
admissible $P$ contains a nonzero vector of the $L_b$ chosen at each step, so no $P$ is missed.
Every partial and complete subspace is recorded by its reduced echelon basis and visited only
once, so every $P$ is produced exactly once.

Let $b_1,\dots,b_m$ be a basis of $B$ and $p_1,\dots,p_m$ a basis of $P$. We store $\Psi$ as
the binary $m\times m$ matrix $(\Psi(b_i,p_k))$. If $P\cap L_b=\{0,y\}$, condition (3) is the
linear equation $\Psi(b,y)=1+y\cdot s_b$ in these $m^2$ entries. If $\dim(P\cap L_b)\ge 2$,
condition (3) says that the equations $\Psi(b,y)=y\cdot s_b$, $y\in P\cap L_b$, do not all
hold. This is not a linear condition. We solve the linear equations first and check the other
conditions on each element of their solution space. If that space is too large to list, we
use a depth-first search instead. It chooses the linear forms $\Psi(b_i,\cdot)$ one at a time
and checks condition (3) for every $b$ in the span of the basis vectors already treated. An additional mode enumerates all dimensions
$0\le\dim B\le M$; the algorithm then returns the complete list of Walsh zero spaces. We used
this mode to compare the algorithm with independent implementations of the generic search
(Section~\ref{sec:exp-ccz}).

\paragraph{Quadratic functions.}
For a quadratic $F$ the data of Theorem~\ref{thm:dual} can be obtained without the Walsh
transform. The space $L_b$ is the radical of the alternating form $b\cdot B(x,y)$, an $n\times
n$ binary matrix. The vector $s_b$ is any solution of the linear system
$s\cdot y=F_b(y)+F_b(0)$, $y\in L_b$. We also split the vector space search in $\NB_F$ over several threads at its first
level. With these improvements the whole test takes about five seconds for the Gold function
in dimension $14$. In dimension $18$ it takes a few minutes, and the dominant cost is the
search for $9$-dimensional subspaces in a set of $87{,}381$ elements.

\paragraph{Cost.}
The running time is governed by the number of subspaces $B$ in the admissible range, the
number of subspaces $P$ for each $B$, and the size of the solution spaces for $\Psi$. These
numbers depend on the function, and we know no useful worst-case bound for them. The
preprocessing stores the sets $\NB_F$ and $L_b$, that is, $O(n2^n)$ words. The search for $P$
also stores the reduced echelon bases of all partial and complete subspaces it visits, and the
algorithm stores the Walsh zero spaces it finds. Neither of these is bounded by $O(n2^n)$ in
general. Both were small for the functions of Section~\ref{sec:experiments}.
Section~\ref{sec:exp-ccz} reports the running times for the functions we tested.

\paragraph{Scope.}
Theorem~\ref{thm:dual} only needs the Walsh supports to be affine. All functions on which we
used it are quadratic, so we have not evaluated the test on other functions with affine
supports. For functions whose Walsh supports are not affine, such as the Kasami function
$x^{57}$ in dimension ten, Proposition~\ref{cor:counting} still applies but
Theorem~\ref{thm:dual} does not. For such functions one has to fall back to the generic
search.

\section{The family \texorpdfstring{$x^3+a^{-1}\Tr(a^3x^9)$}{x\^3+a\^-1Tr(a\^3x\^9)}}\label{sec:family}

In this section $n=2m$ and $q=2^m$. We write $C=\{x^3:x\in\Ftn^*\}$ for the subgroup of cubes,
which has index three since $n$ is even. Let $\alpha$ be a primitive element of $\Ftn$. We study
\[
F_a(x)=x^3+a^{-1}\Tr(a^3x^9),\qquad a\in\Ftn^*,
\]
the family of Budaghyan, Carlet and Leander \cite{DBLP:journals/ffa/BudaghyanCL09}
(family F4 above). Göloğlu and Pavlů proved that $F_a$ is not CCZ-equivalent to a permutation
when $m$ is even \cite[Theorem~3]{DBLP:journals/ccds/GologluP21}. Their proof shows that
$\NB_{F_a}\cup\{0\}$ contains no $m$-dimensional subspace, using that for even $m$ the set
$C\cup\{0\}$ contains no subspace of dimension $m$
\cite[Lemma~13]{DBLP:journals/ffa/GologluL21}. We study the case of odd $m$, where this
fails.

\begin{lemma}\label{lem:Fa-classes}
For $c\in\Ftn^*$ we have $F_a(cx)=c^3F_{ac^3}(x)$ and $F_a(x)^2=F_{a^2}(x^2)$. Hence $F_a$
and $F_{a'}$ are linearly equivalent whenever $a'\in aC$ or $a'=a^2$, and every member of the
family is linearly equivalent to $F_1$ or to $F_\alpha$.
\end{lemma}

\begin{proof}
$c^3F_{ac^3}(x)=c^3x^3+c^3(ac^3)^{-1}\Tr(a^3c^9x^9)=F_a(cx)$, and
$F_a(x)^2=x^6+a^{-2}\Tr(a^6x^{18})=F_{a^2}(x^2)$ since $\Tr(y)^2=\Tr(y)$. The cosets of $C$ in
$\Ftn^*$ are $C$, $\alpha C$ and $\alpha^2C$, and $a\mapsto a^2$ maps $\alpha C$ onto $\alpha^2C$.
\end{proof}

The components of $F_a$ are
$\Tr(\beta F_a(x))=\Tr(\beta x^3)+\Tr(\beta a^{-1})\Tr(a^3x^9)$. On the hyperplane
$H_a=\{\beta:\Tr(\beta a^{-1})=0\}$ they coincide with the components of $x^3$, which are
bent exactly when $\beta\notin C$ \cite[Lemma~1]{DBLP:journals/ccds/GologluP21}.
For odd $m$ we have $\gcd(3,q-1)=1$, so $\F_q^*\subseteq C$, and $C\cup\{0\}$ is the union of
the $(q+1)/3$ lines $c\F_q$, $c\in C$, of the Desarguesian spread $\{c\F_q: c\in\Ftn^*\}$ of
$\Ftn$.

\begin{proposition}\label{prop:Fa-line}
Let $m$ be odd and $a\in C$. Then the $m$-dimensional subspace $a\F_q$ is contained in
$\NB_{F_a}\cup\{0\}$. In particular the necessary condition of
Proposition~\ref{cor:counting} does not exclude $F_a$ by dimension alone.
\end{proposition}

\begin{proof}
Let $\beta=at$ with $t\in\F_q^*$. Then $\beta\in C$ because $a\in C$ and $t\in\F_q^*\subseteq C$,
and $\Tr(\beta a^{-1})=\Tr(t)=\Tr_{\F_q}(t+t^q)=0$, so $\beta\in H_a$. The component
$\Tr(\beta F_a)$ is therefore the component $\Tr(\beta x^3)$ of the Gold function, which is not
bent since $\beta\in C$.
\end{proof}

Here $\NB_{F_a}$ is taken with respect to the trace form, i.e.\ in the field coordinates
$\beta$ of the components $x\mapsto\Tr(\beta F(x))$. The computations below are carried out in
bit coordinates and translated to field coordinates through the dual basis of the trace form.

\subsection{Computations for \texorpdfstring{$n\in\{6,10,14,18,22\}$}{n in \{6,10,14,18,22\}}}

By Lemma~\ref{lem:Fa-classes} it suffices to consider $a\in\{1,\alpha\}$, where $1\in C$
and $\alpha\notin C$. For $n\le 18$ we nevertheless computed $a=\alpha^2$ as well, with the
same results as for $a=\alpha$. Table~\ref{tab:Fa} lists, for each case, the $m$-dimensional
subspaces of $\NB_{F_a}\cup\{0\}$. In every case there is none of dimension $m+1$. So $M\le m$, with
equality whenever an $m$-dimensional subspace exists. By Proposition~\ref{cor:counting} a
permutation then requires two $m$-dimensional subspaces $B_1,B_2$ with $B_1\cap B_2=\{0\}$. If there is at most one $m$-dimensional subspace, this
excludes a permutation. For $n=22$
this follows from the uniqueness of the $m$-dimensional subspace, since a subspace of
dimension $m+1$ contains $2^{m+1}-1$ subspaces of dimension $m$. For $n=18$ the set
$\NB_{F_a}$ has $87{,}381$ elements. The search for subspaces of dimension nine and ten was
split over 128 threads and took about five minutes per value of $a$.

For $n=22$ the set $\NB_{F_a}$ has $1{,}398{,}101$ elements, and the plain search is no longer
feasible. We therefore used the symmetries of the family. The two identities of
Lemma~\ref{lem:Fa-classes} give $F_a(ax^4)=a^3F_a(x)^4$, and moreover $F_1(x^2)=F_1(x)^2$. So,
in bit coordinates, $\NB_{F_a}$ is invariant under the transpose of the linear map $y\mapsto
a^3y^4$, which has order $11$. For $a=1$, $\NB_{F_1}$ is also invariant under the transpose of
$y\mapsto y^2$, which has order $22$. The subspaces are then searched only through the smallest
element of each orbit, removing the orbits already treated, and the results are closed under
the group. These searches are independent, so they can be distributed over several machines.
For $a=1$ the search over the $63{,}644$ orbit representatives took about $63$ hours on $128$
threads. For $a=\alpha$ the $127{,}101$ representatives were split over up to three servers with
$200$ threads each, and every finished representative was saved. This search took about three
days.

The orbit-pruned search works as follows. Let $T$ be the linear map that generates the group
$G=\langle T\rangle$, and let $Z=\NB_{F_a}$ in bit coordinates, so that $T(Z)=Z$.
\begin{enumerate}
\item List the smallest elements $r_1<r_2<\dots<r_k$ of the $G$-orbits in $Z$, as integers.
\item For each $i$, let $Z_i$ be $Z$ minus the orbits of $r_1,\dots,r_{i-1}$. Find all
$d$-dimensional subspaces $V$ with $r_i\in V\subseteq Z_i\cup\{0\}$. Their images in the quotient
$\Fn/\langle r_i\rangle$ are the $(d-1)$-dimensional subspaces contained in
$Q_i=\{0\}\cup\{w+\langle r_i\rangle: w\in Z_i,\ w+r_i\in Z_i\}$, which we find with the plain
search.
\item Return the union of the $G$-orbits of all subspaces found, deduplicated by reduced
echelon bases.
\end{enumerate}
No subspace is missed. Let $V\subseteq Z\cup\{0\}$ have dimension $d$, and let $i$ be the
smallest index such that $V$ meets the orbit of $r_i$, say in $g(r_i)$ with $g\in G$. Then
$g^{-1}(V)$ contains $r_i$, and it meets no orbit of $r_j$ with $j<i$, because these orbits
are $G$-invariant. So $g^{-1}(V)$ is found in step~2 for this $i$, and $V$ in step~3.
Subspaces with a nontrivial stabilizer need no special treatment. They may be found several
times, and step~3 removes the duplicates. The steps for different $i$ are independent. For
$a=\alpha$ they were run in twelve lanes ($i$ modulo $12$), each with a checkpoint file of
finished indices. A final check confirms that every index from $1$ to $127{,}101$ is present. We also checked the implementation against the plain search with the same groups
for $n=14$ (dimensions $5$ to $8$, with up to $5{,}348$ subspaces) and for $n=18$ (dimensions
$9$ and $10$), for $a\in\{1,\alpha,\alpha^2\}$. The two lists of subspaces coincide in every
case (Section~\ref{sec:exp-family}).

We also timed a sample of $800$ representatives for each value of $a$ on the same server, at
the same relative positions in the two lists of representatives. A single search costs the
same for both values: the ratio of the mean CPU times is $0.97$. The total cost is therefore
proportional to the number of representatives. It is about $13{,}000$ thread-hours for $a=1$
and $25{,}000$ for $a=\alpha$, measured as CPU time of the hardware threads of a fully loaded
server. The factor of two comes from the symmetry group, whose order is $11$ instead of $22$.

\begin{table}[t]
\caption{Number (\#) of $m$-dimensional subspaces of $\NB_{F_a}\cup\{0\}$ for
$F_a(x)=x^3+a^{-1}\Tr(a^3x^9)$, $n=2m$, $m$ odd. ``Lines'' means that every such subspace is a line $c\F_q$ of the
Desarguesian spread; ``$a\F_q$'' that one of them is $a\F_q$. The necessary condition of
Proposition~\ref{cor:counting} (column ``condition'') is inconclusive when two of the
subspaces intersect trivially.
In every case $F_a$ is not CCZ-equivalent to a permutation.}\label{tab:Fa}
\begin{tabular*}{\textwidth}{@{\extracolsep\fill}cclccccl}
\toprule
$n$ & $m$ & $a$ & \# & lines & $a\F_q$ & condition & decided by \\
\midrule
6 & 3 & $1\in C$ & 3 & yes & yes & inconcl. & Thm.~\ref{thm:dual} \\
6 & 3 & $\alpha,\alpha^2\notin C$ & 0 & -- & -- & excludes & Prop.~\ref{cor:counting} \\
10 & 5 & $1\in C$ & 1 & yes & yes & excludes & Prop.~\ref{cor:counting} \\
10 & 5 & $\alpha,\alpha^2\notin C$ & 1 & yes & no & excludes & Prop.~\ref{cor:counting} \\
14 & 7 & $1\in C$ & 1 & yes & yes & excludes & Prop.~\ref{cor:counting} \\
14 & 7 & $\alpha,\alpha^2\notin C$ & 1 & yes & no & excludes & Prop.~\ref{cor:counting} \\
18 & 9 & $1\in C$ & 3 & yes & yes & inconcl. & Thm.~\ref{thm:dual} \\
18 & 9 & $\alpha,\alpha^2\notin C$ & 0 & -- & -- & excludes & Prop.~\ref{cor:counting} \\
22 & 11 & $1\in C$ & 1 & yes & yes & excludes & Prop.~\ref{cor:counting} \\
22 & 11 & $\alpha,\alpha^2\notin C$ & 1 & yes & no & excludes & Prop.~\ref{cor:counting} \\
\botrule
\end{tabular*}
\end{table}

For $n=18$ and $a=1$ the three subspaces intersect pairwise trivially, so they span
$\F_2^{18}$ in pairs and the necessary condition is satisfied. Algorithm~\ref{alg:dual} shows
that none of them is the projection of a Walsh zero space. Once the subspaces are known, this
takes two seconds. The failure occurs in the last step. For each of the three subspaces $B$,
every $L_b$ with $b\in B\setminus\{0\}$ has dimension two, and there are exactly three
admissible subspaces $P$. Each of them meets every $L_b$ in a single nonzero vector. So
condition (3) of Theorem~\ref{thm:dual} is a system of $511$ linear equations in the $81$
entries of $\Psi$, and this system has no solution. For each of the nine systems the repository contains a witness of
infeasibility: a set of $7$ to $23$ of the equations whose left-hand sides sum to zero and whose
right-hand sides sum to one. It also contains the bases of the three subspaces. The same
happens for $n=6$ and $a=1$: each of the three candidate projections has three admissible $P$,
and each resulting system of $7$ equations in the $9$ entries of $\Psi$ has no solution. Together with the theorem of Göloğlu and Pavlů
this gives the following statement. It is a computer-assisted result for
$n\in\{6,10,14,18,22\}$.

\begin{corollary}\label{cor:Fa-20}
For every even $n\le 24$ and every $a\in\F_{2^n}^*$, the function $x^3+a^{-1}\Tr(a^3x^9)$ is
not CCZ-equivalent to a permutation.
\end{corollary}

\begin{proof}
For $n\in\{4,8,12,16,20,24\}$ this is \cite[Theorem~3]{DBLP:journals/ccds/GologluP21}; for
$n=2$ there is no APN permutation of $\F_2^2$. For $n\in\{6,10,14,18,22\}$ it follows from
Lemma~\ref{lem:Fa-classes} and Table~\ref{tab:Fa}.
\end{proof}

For $n=10$ and $n=14$ the necessary condition suffices. These cases are therefore within
reach of the method of \cite{DBLP:journals/ccds/GologluP21}, which reports computations for
$n\le 12$ and random members of the known families up to $n=18$. For $n=18$ and $a\in C$, on
the other hand, the necessary condition is inconclusive and no theorem applies. To our
knowledge this case is new, and it is decided by Theorem~\ref{thm:dual}.

\paragraph{Subspaces of the cubes.}
Behind Table~\ref{tab:Fa} is the structure of the large subspaces of $C\cup\{0\}$. For $m=5$
and $m=7$, every subspace of $C\cup\{0\}$ of dimension $m$ or $m-1$ lies in a single spread
line. There are $11$ and $341$ such subspaces for $n=10$, and $43$ and $5{,}461$ for $n=14$.
For $m=3$, in contrast, only $21$ of the $28$ two-dimensional subspaces do. If this holds for
all odd $m\ge 5$, the proof of \cite[Theorem~3]{DBLP:journals/ccds/GologluP21} can be followed
with $V=W\cap H_a$ contained in one spread line. For $\dim V=m$ one obtains exactly the line of
Proposition~\ref{prop:Fa-line}. For $\dim V=m-1$ the character-sum argument of
\cite{DBLP:journals/ccds/GologluP21} no longer gives a contradiction, by the following lemma.
Here $V^\perp$ is taken with respect to the trace form.

\begin{lemma}\label{lem:Vperp}
Let $m$ be odd, $c\in C$, and let $V$ be an $(m-1)$-dimensional subspace of the spread line
$c\F_q$. Then $|V^\perp\cap C|=(4q-5)/3$.
\end{lemma}

\begin{proof}
Write $V=cH$ with $H$ a hyperplane of $\F_q$, and let $T$ be the trace from $\Ftn$ to $\F_q$.
For $\beta\in\Ftn$ and $h\in\F_q$ we have $\Tr(\beta ch)=\Tr_{\F_q}(h\,T(\beta c))$. So
$\beta\in V^\perp$ if and only if $T(\beta c)\in\{0,s\}$, where $\{0,s\}$ is the orthogonal of
$H$ in $\F_q$. Put $\gamma=\beta c$; since $c\in C$, $\beta\in C$ if and only if
$\gamma\in C$. The kernel of $T$ is $\F_q$, and $\F_q^*\subseteq C$, which gives $q-1$
elements with $T(\gamma)=0$. Multiplication by $t\in\F_q^*$ preserves $C$ and multiplies $T$
by $t$. So all nonzero values of $T$ are taken equally often on $C$, namely
$(|C|-(q-1))/(q-1)=(q+1)/3-1=(q-2)/3$ times. The total is $q-1+(q-2)/3=(4q-5)/3$.
\end{proof}

The data suggest the following.

\begin{conjecture}\label{conj:Fa}
Let $m\ge 5$ be odd and $n=2m$. If $3\nmid m$, then $\NB_{F_a}\cup\{0\}$ contains exactly one
subspace of dimension $m$, a line of the Desarguesian spread, equal to $a\F_q$ if and only if
$a\in C$. If $3\mid m$, it contains three such subspaces if $a\in C$ and none otherwise, and
none of them is the projection of a Walsh zero space of $F_a$. In particular $F_a$ is not
CCZ-equivalent to a permutation for any even $n$.
\end{conjecture}

The case $3\mid m$ of the conjecture rests on a single value in its range, $m=9$. The case
$m=3$ lies outside the range but behaves in the same way. The case $3\nmid m$ rests on
$m=5,7,11$. The case $m=11$ ($n=22$) was computed after the conjecture was
formed, and agrees with it for both $a\in C$ and $a\notin C$. Lemma~\ref{lem:Vperp} was also checked numerically for $n=10$ and $n=14$.

\section{Experiments}\label{sec:experiments}

\subsection{Setup and reproducibility}

The core routines are written in C and called from Python. They compute difference tables
and the cost~\eqref{eq:cost}, and implement incremental local search, vector space search,
switching kernels, and an APN test for quadratic functions from their bilinear form. The test
of Section~\ref{sec:ccz}, the invariants and the experiment drivers are written in Python with
NumPy. Unless stated otherwise, experiments were run on a laptop with an Apple M5 Pro
processor (18 cores, 24\,GB of memory). The randomized tree-search campaign of
Section~\ref{sec:exp-new} and experiments E7, E10 and E12 were run on servers. Each server has two AMD EPYC
7763 processors ($128$ physical cores, $256$ hardware threads and $1$\,TB of memory), and we ran at low priority, in containers limited to $200$ CPUs. The number of worker processes
is given with each experiment.

The accompanying repository (\url{https://github.com/peacker/ccz-projection-test-apn})
contains one script per experiment (E1--E12 below) and the data files they produce. Running a
script recomputes its results. A separate script compares every number reported in this
section with the stored results; it does not recompute them. The repository gives the field
representations (primitive polynomials) and the bit order. The random seeds are fixed in the
scripts.

The checks validate different things.
\begin{itemize}
\item APN property: every APN function found by our search programs (dimensions six to ten)
was checked by computing its full difference table, independently of the method that produced
it. In larger dimensions only the functions of known families are used.
\item Permutations: every permutation produced by the CCZ test was checked to be a bijection
and APN.
\item Equivalences: every EA-equivalence we claim comes with an explicit certificate
$(L_1,L_2)$, and a separate routine checks that $L_1,L_2$ are invertible and that
$G+L_1\circ F\circ L_2$ is affine.
\item Walsh zero spaces: the projection test and sboxU return the same lists for the $14$
functions of Table~\ref{tab:ccz-validation}.
\item Completeness of exhaustive searches: this cannot be checked from the output. We rely on
cross-checks with independent implementations. sboxU checks the Walsh zero spaces, and the
plain search checks the symmetry-reduced search (E10). The known classification in dimension
six checks the switching closure (E5).
\end{itemize}
The code, the experiments and a first draft of the manuscript were prepared with the
assistance of a large language model (Claude, Anthropic). The author takes full responsibility
for the content. The checks listed above are run by the scripts of the repository.

\subsection{The inverse function and structured permutations (E1, E2)}\label{sec:exp-inverse}

For $n\in\{4,6,8,10\}$ we verified Proposition~\ref{prop:inverse} directly. For every
$a\neq0$ the difference table of $I$ has exactly one entry equal to $4$, and its four
solutions form the line $a\F_4$. For $n=8$ we scanned all $32{,}640$ transpositions of $I$.
None of them decreases the cost. Exactly $255$ leave it unchanged: those involving position $0$,
each of which replaces $6$ entries equal to $4$ by two entries equal to $6$. All others
increase the cost by at least $222$. The $255$ permutations reached by the cost-preserving
moves cannot be improved by a further transposition either. We ran simulated annealing from
$I$ for $2\cdot10^7$ moves per run, without restarts. The temperature decreases geometrically
from $T_0$ to $T_1$, with $(T_0,T_1)\in\{(2,0.1),(20,0.5),(60,1)\}$, and a move that increases
the cost by $d$ is accepted with probability $e^{-d/T}$. We used plain transpositions and
involution-preserving moves, with three seeds per setting. An involution-preserving move picks
two points $x\neq y$; if they form a $2$-cycle it splits it into two fixed points, and
otherwise it makes $\{x,y\}$ a $2$-cycle and pairs their former partners. No run found a
permutation of cost below $255$. With the same budget and $(T_0,T_1)=(8,0.2)$, $18$ annealing
runs started from random permutations end at costs between $4{,}776$ and $4{,}926$. The seeds
are fixed in the script.

The families of Section~\ref{sec:constructions} were enumerated exhaustively, with the
results of Table~\ref{tab:constructions}. The butterfly enumeration for $n=10$ (190,464
permutations) takes about three minutes.

\subsection{The projection test compared with a generic search (E3, E12)}\label{sec:exp-ccz}

We compared the projection test (Algorithm~\ref{alg:dual}) with the library sboxU
\cite{sboxU}. sboxU implements the generic vector space search of
\cite{DBLP:conf/asiacrypt/BonnetainPT19}, and was used to test CCZ-equivalence to permutations
in \cite{DBLP:journals/tit/BeierleL22,DBLP:journals/ccds/BudaghyanIK23,DBLP:journals/corr/abs-2508-04644}.
Both programs ran on the same otherwise idle server (two AMD EPYC 7763 processors), with sboxU
at commit \texttt{bedb6d1}. Library loading is not included in the times. We measured two
tasks. The first is listing all Walsh zero spaces, with one thread for both programs: the
mode of Algorithm~\ref{alg:dual} that enumerates every dimension $0\le\dim B\le M$, and the
function \texttt{get\_WalshZeroesSpaces} of sboxU. The second is deciding whether the class
contains a permutation: Algorithm~\ref{alg:dual} with one thread, and the function
\texttt{enumerate\_permutations\_in\_ccz\_class} of sboxU with all available threads.

Table~\ref{tab:ccz-validation} shows the results. Each time is a single run, so the
benchmark is illustrative and we make no statistical claim about the ratios. The two programs
find the same Walsh zero spaces for every function, and both find permutations in the class of
the Kim function. For $n\le 8$ both take a fraction of a second, and sboxU is faster for
$n=6$. From $n=10$ on the projection test is faster. Listing all Walsh zero spaces is the
cleanest comparison of the two enumeration methods. The projection test takes $5$ to $18$ times
less time in dimension ten, and $39$ to $55$ times less in dimension twelve.

For deciding, the column ``by'' of Table~\ref{tab:ccz-validation} gives a sufficient reason
for the answer. For $M$ and $\Psi$ it is also the step at which our implementation stops. For
the two functions marked $1$, a single candidate projection already excludes a permutation,
because $M=n/2$; our implementation does not use this shortcut and still runs the lifting
step. In dimension twelve both functions have $M=4<6$, so the dimension bound of
Proposition~\ref{cor:counting} settles them before any projection is examined. The large
ratio in that row therefore measures this bound, which sboxU does not apply, and not
Theorem~\ref{thm:dual}. A generic search preceded by the same bound would also be fast there.
In dimension ten three of the functions need the lifting step of Theorem~\ref{thm:dual}: $x^3$,
$x^9$ and the Beierle--Leander function. For them deciding takes $0.16$ to $0.24$ seconds,
against $5.5$ to $7.6$ seconds for sboxU, that is, $23$ to $47$ times less.

Three further points limit this comparison. First, sboxU used all threads and the projection
test one. Second, the function of sboxU computes all permutations in the class, not only whether one
exists; here, however, there are none except for the Kim function. Third, the projection test
applies only to functions with affine Walsh supports. For the Edel--Pott function in dimension
six and the Kasami function in dimension eight, whose supports are not affine, we used our own
generic search (E3). It finds $25$ and $1$ Walsh zero spaces, respectively.

\begin{table}[t]
\caption{The projection test (Alg.~\ref{alg:dual}) and the generic search of sboxU
\cite{sboxU} on the same machine (E12). ``Spaces'' is the number of Walsh zero spaces and
``perm.'' whether the CCZ-class contains a permutation; times are in seconds. ``List'': all
Walsh zero spaces, one thread for both; ``decide'': Alg.~\ref{alg:dual} with one thread, sboxU
(\texttt{enumerate\_permutations\_in\_ccz\_class}) with all threads. ``By'' is a sufficient
reason for the answer: $M$ the dimension bound, 1 a single candidate projection (with $M=n/2$),
$\Psi$ the lifting step of Theorem~\ref{thm:dual}, perm.\ a permutation found.}\label{tab:ccz-validation}
\begin{tabular*}{\textwidth}{@{\extracolsep\fill}llrcrrrrc}
\toprule
$n$ & Function & Spaces & perm. & \multicolumn{2}{c}{List} & \multicolumn{3}{c}{Decide} \\
\cmidrule{5-6}\cmidrule{7-9}
 & & & & Alg.~\ref{alg:dual} & sboxU & Alg.~\ref{alg:dual} & sboxU & by \\
\midrule
6 & $x^3$ & 190 & no & 0.030 & 0.004 & 0.007 & 0.022 & $\Psi$ \\
6 & Kim function & 222 & yes & 0.036 & 0.006 & 0.009 & 0.034 & perm. \\
8 & $x^3$ & 256 & no & 0.10 & 0.11 & 0.025 & 0.064 & $M$ \\
8 & $x^3+\Tr(x^9)$ & 256 & no & 0.11 & 0.12 & 0.025 & 0.065 & $M$ \\
8 & $x^9$ & 256 & no & 0.10 & 0.12 & 0.024 & 0.064 & $M$ \\
10 & $x^3$ & 1{,}024 & no & 3.37 & 19.9 & 0.24 & 5.47 & $\Psi$ \\
10 & $x^3+\Tr(x^9)$ & 1{,}744 & no & 3.37 & 17.5 & 0.15 & 6.90 & 1 \\
10 & $x^9$ & 1{,}024 & no & 3.46 & 19.8 & 0.24 & 5.54 & $\Psi$ \\
10 & $x^3+\alpha^{-1}\Tr(\alpha^3x^9)$ & 1{,}624 & no & 3.21 & 17.6 & 0.15 & 6.56 & 1 \\
10 & Beierle--Leander \#3 & 1{,}644 & no & 2.78 & 24.3 & 0.16 & 7.59 & $\Psi$ \\
10 & F3 (one class) & 1{,}024 & no & 2.26 & 40.8 & 0.14 & 8.63 & $M$ \\
10 & F12 (one class) & 1{,}024 & no & 2.55 & 34.5 & 0.14 & 7.15 & $M$ \\
12 & $x^3$ & 12{,}286 & no & 81.6 & 3{,}200 & 0.63 & 744 & $M$ \\
12 & $x^3+\Tr(x^9)$ & 8{,}038 & no & 64.1 & 3{,}507 & 0.82 & 727 & $M$ \\
\botrule
\end{tabular*}
\end{table}

\subsection{Known quadratic APN classes in dimension ten (E4, E11)}\label{sec:exp-known}

We instantiated the families F3, F4, F12 and F13 of
\cite[Table~3]{DBLP:journals/ccds/Kaleyski21} in dimension ten, together with the Gold
functions. For F4 we took $x^3+a^{-1}\Tr(a^3x^9)$ with $a\in\F_4^*$ and with a primitive $a$.
For F3 and F12 we took $300$ random parameter choices each. For F13 we took all functions of
the shape $x^3+a(x^{2^i+1})^{2^k}+bx^{3\cdot 2^5}+c(x^{2^{i+5}+2^5})^{2^k}$ with
$a,b,c\in\F_4$ and all $i,k$; these $5{,}760$ functions include the family of
\cite{DBLP:journals/tit/BudaghyanHK20}. Of the $6{,}366$ functions, $1{,}266$ are APN. They have
$17$ distinct ortho-derivative invariants. We added the five functions of
\cite{DBLP:journals/tit/BeierleL22,BeierleL21data}. Two of them have the same invariants as
F13-type functions found above (with $M=5$), so this gives twenty invariants. The two
coincidences are in fact EA-equivalences (E11, see below).

Every distinct APN instance is certified EA-equivalent to the representative of its group by
the explicit EA-equivalence test described below. So the twenty groups are exactly twenty
classes. None of them is CCZ-equivalent to a permutation. For $13$ of them the set of
non-bent components contains no subspace of dimension $5$. For these, the dimension bound of
Proposition~\ref{cor:counting} already excludes a permutation. Our values of $M$ for the
families F3, F4, F12 and the Budaghyan--Helleseth--Kaleyski functions agree with those
reported in \cite[Table~3]{DBLP:journals/ccds/GologluP21} (families 6, 7, 12 and 10 there).
For the remaining seven classes, the non-bent components contain between one and eleven
subspaces of dimension $5$. Algorithm~\ref{alg:dual} shows that none of them is the
projection of a Walsh zero space. For four of these classes there is a single candidate, so
the spanning condition of Proposition~\ref{cor:counting} already excludes a permutation. For the
other three the necessary condition is satisfied: the two Gold classes and the class with two
candidate projections. There the
candidate projections intersect pairwise trivially and therefore span $\F_2^{10}$, so
Proposition~\ref{cor:counting} is inconclusive and the projection test is needed. For the Gold
functions the conclusion also follows from the theorem of \cite{DBLP:journals/ffa/GologluL21}.
For the other class it was obtained in \cite{DBLP:journals/tit/BeierleL22} with sboxU.
Algorithm~\ref{alg:dual} decides these cases in $0.15$ to $0.25$ seconds, while sboxU takes $5$
to $8$ seconds on the same machine (Section~\ref{sec:exp-ccz}).

To make the conclusion independent of the invariant, we applied the test to every distinct APN
function of the experiment, $402$ functions in total. None is CCZ-equivalent to a permutation,
and $256$ of them are already excluded by the dimension bound of
Proposition~\ref{cor:counting} ($M<5$). For the families
known at the time, this confirms the result of Göloğlu and Pavlů
\cite{DBLP:journals/ccds/GologluP21} by an independent computation. The F13-type classes with
parameters in $\F_4$ that have $M=5$ contain instances of \cite{DBLP:journals/tit/BeierleL22},
which were tested there.

Experiment E11 treats the sources of quadratic APN functions in dimension ten that the list
above does not cover. We group their APN instances by the ortho-derivative invariant. Then we
prove every coincidence by an explicit EA-equivalence $G=L_1\circ F\circ L_2+A$. In that case
the ortho-derivatives satisfy $\pi_G=(L_1^{\top})^{-1}\circ\pi_F\circ L_2$, so the equivalence
can be recovered through the ortho-derivative, as in \cite{DBLP:journals/tit/CanteautCP22}. We
use a plain depth-first search over the images of a basis under $L_2$, pruned by point
invariants of $\pi_F$ and $\pi_G$. A candidate is accepted only if $G+L_1\circ F\circ L_2$ is
affine, which we verify directly. (A general algorithm for this problem is given by Kaleyski
and Sunde \cite{KaleyskiS26}.) By \cite{DBLP:journals/jalc/Yoshiara12}, EA- and
CCZ-equivalence coincide for quadratic APN functions. The search is exhaustive. It tries every image of a basis under $L_2$ that is compatible
with the relation $\pi_G=(L_1^{\top})^{-1}\circ\pi_F\circ L_2$, and the point invariants it uses
for pruning are preserved by that relation, so no equivalence is lost. It takes at most a few
seconds per pair. The certificates of E4, E5 and E11 are stored in the repository, together
with a short script that checks them without the search. Each of the $373$ APN instances of E11 is certified to be EA-equivalent to the
representative of its class.
\begin{itemize}
\item Zheng et al.\ \cite{DBLP:journals/tit/ZhengKLPT22}. We used the three instances given
explicitly in the preprint version of their paper (arXiv:2101.11535v1, Table~III): the
function of their family F15 and the two sporadic instances. We took them for every primitive
element of $\F_{2^{10}}$ up to Frobenius. We also took $200$ random members of the family of
their Theorem~3.3. The first two instances are APN for every primitive element, the third only
for two of the $60$ primitive elements up to Frobenius. All of them are EA-equivalent to known
functions. The first two lie in the F13-type classes that contain the fourth and the second
instance of \cite{DBLP:journals/tit/BeierleL22}, respectively. The third is equivalent to the
third instance of \cite{DBLP:journals/tit/BeierleL22}. The APN members of the family lie in
the classes of $x^3$, F3 and F13. We could not access a list of the instances of the journal
version, whose abstract mentions two instances of an infinite family.
\item Göloğlu's biprojective families $\mathcal{F}_1$ and $\mathcal{F}_2$
\cite{DBLP:journals/tit/Gologlu22} for all $q$, and the first family of Calderini, Li and Villa
\cite[Theorem~3]{DBLP:journals/ffa/CalderiniLV23} for all parameters. $\mathcal{F}_1$ and the
latter family give two of the F13-type classes. $\mathcal{F}_2$ gives the two F13-type classes
that contain the second and fourth instance of \cite{DBLP:journals/tit/BeierleL22}. These are
the $\varphi(5)/2=2$ classes of each of $\mathcal{F}_1$ and $\mathcal{F}_2$ counted in
\cite[Theorem~5.1]{DBLP:journals/combtheory/GologluK25}. Random members of Carlet's family give
the two classes of F3.
\item Budaghyan, Ivkovic and Kaleyski \cite[Table~2]{DBLP:journals/ccds/BudaghyanIK23}. All six
functions, instantiated with either primitive element of $\F_4$, are APN. Their
ortho-derivatives have exactly the differential spectra printed there. Five of them lie outside
the twenty classes above and have pairwise different invariants, so they are five further
classes. The second function of their table, as printed, is EA-equivalent to the function
$x^3+\beta x^{36}$, $\beta$ primitive in $\F_4$, of Edel, Kyureghyan and Pott
\cite{DBLP:journals/tit/EdelKP06}. This is the F13-type class with $(a,b,c)=(\beta,0,0)$,
$i=3$, $k=2$ above.
\item The function of Li, Zhou, Li and Qu (\cite{DBLP:journals/tit/LiZLQ22}; Theorem~5 of the preprint arXiv:2105.08464v1),
and all eleven APN members of the second family of Calderini, Li and Villa
\cite[Theorem~4]{DBLP:journals/ffa/CalderiniLV23}, which contains it. They are EA-equivalent
to the first function of \cite[Table~2]{DBLP:journals/ccds/BudaghyanIK23}.
\item Yu et al.\ \cite{YuLIK25}. The representative
$x^{513}+x^{192}+x^{96}+x^{48}+x^{33}+x^{18}+x^{12}$ of the class of $x^3+\Tr(x^9)$ given there
is EA-equivalent to $x^3+\Tr(x^9)$, as stated in \cite{YuLIK25}. The other two classes of that
classification are those of $x^3$ and $x^9$.
\end{itemize}
For each of the five further classes, the non-bent components contain no subspace of
dimension five ($M=4$). So Proposition~\ref{cor:counting} excludes a permutation. This agrees
with the check made with sboxU in \cite{DBLP:journals/ccds/BudaghyanIK23}, and for these
classes our test is an independent confirmation. In total, the instances we examined give $25$
classes of quadratic APN functions in dimension ten (Table~\ref{tab:known10}). None of them is
CCZ-equivalent to a permutation.

The count $25$ is exact for the instances we considered, but it is not a classification of
the families. For F4 we took representatives of all parameters up to the equivalences of
Lemma~\ref{lem:Fa-classes}. We took all parameters for F13 with parameters in $\F_4$, for the
biprojective families of Göloğlu and for the two families of Calderini, Li and Villa. We also
took all published sporadic instances we could obtain. For F3 and F12 we took $300$ random parameter
choices each, for Carlet's family $50$, and for the family of Zheng et al.\ $200$. For Zheng et
al.\ we used the instances of the preprint version. A class of these sampled families that our
samples missed would be a known class that we did not test.

\begin{table}[t]
\caption{The $25$ classes of quadratic APN functions in dimension ten represented by the
instances examined in E4 and E11. $M$ is
the largest dimension of a subspace of $\NB_F\cup\{0\}$ (there are $341$ non-bent components
in every case). The last column says what excludes a permutation: ``$M<5$'' is the dimension
bound of Proposition~\ref{cor:counting}; ``one projection'' means that $\NB_F\cup\{0\}$
contains a single $5$-dimensional subspace, so the spanning condition of
Proposition~\ref{cor:counting} fails; ``Thm.~\ref{thm:dual}'' means that the candidate
projections span $\F_2^{10}$ in pairs and the projection test is needed. No class is
CCZ-equivalent to a permutation.}\label{tab:known10}
\begin{tabular*}{\textwidth}{@{\extracolsep\fill}lccl}
\toprule
Family & Classes & $M$ & Reason \\
\midrule
Gold $x^3$, $x^9$ & 2 & 5 & Thm.~\ref{thm:dual} (11 projections) \\
F4: $x^3+a^{-1}\Tr(a^3x^9)$ \cite{DBLP:journals/ffa/BudaghyanCL09} & 2 & 5 & one projection \\
F3 \cite{DBLP:journals/tit/BudaghyanC08} & 2 & 4 & $M<5$ \\
F12 \cite{DBLP:journals/dcc/Taniguchi19} & 6 & 3, 4 & $M<5$ \\
F13-type, parameters in $\F_4$ \cite{DBLP:journals/tit/BudaghyanHK20} & 5 & 3, 4, 5 & $M<5$ (3), one projection (2) \\
Beierle--Leander \cite{DBLP:journals/tit/BeierleL22} & 3 & 3, 4, 5 & $M<5$ (2), Thm.~\ref{thm:dual} (1; 2 projections) \\
Budaghyan et al.\ \cite{DBLP:journals/ccds/BudaghyanIK23}, \cite{DBLP:journals/tit/LiZLQ22} & 5 & 4 & $M<5$ \\
\midrule
Total & 25 & & $M<5$: 18; one projection: 4; Thm.~\ref{thm:dual}: 3 \\
\botrule
\end{tabular*}
\end{table}

\subsection{The family \texorpdfstring{$x^3+a^{-1}\Tr(a^3x^9)$}{x\^3+a\^-1Tr(a\^3x\^9)} (E9, E10)}\label{sec:exp-family}

The computations of Section~\ref{sec:family} are experiments E9 and E10. E9 covers
$n\in\{6,10,14\}$ and the subspaces of the cubes, on the laptop. E10 covers $n=18$ and $n=22$,
on the servers. For each $n$ and each $a$, they list the subspaces of $\NB_{F_a}\cup\{0\}$ of
dimensions $m$ and $m+1$, translate them to field coordinates through the dual basis of the
trace form, and run Algorithm~\ref{alg:dual}. For $n=18$ the computation of $\NB_{F_a}$ takes
about one minute and the parallel subspace search about five minutes. The final step of
Algorithm~\ref{alg:dual} takes two seconds. The same subspace search on a single core took four
hours. The results are those of Table~\ref{tab:Fa}. For $n=22$, E10 uses the symmetry
reduction of Section~\ref{sec:family}. For $a=1$ it ran in one go, in about $63$ hours on $128$
threads; computing $\NB_{F_1}$ takes $20$ minutes. For $a=\alpha$ the orbit representatives
were split into $12$ lanes that ran on up to three servers. Every finished representative was
saved with its index and the subspaces found, so the work could be moved between servers and
survive reboots. The repository contains these lane files, the merged result, and a check that
every index from $1$ to $127{,}101$ is present. A list of SHA-256 hashes of all result files
fixes their content.

Two further computations support these results. First, we compared the symmetry-reduced
search with the plain search, with the same groups (E10, symmetry check). For $n=14$ we used
dimensions $5$ to $8$ and found up to $5{,}348$ subspaces, in about a minute on the laptop. For
$n=18$ we used dimensions $9$ and $10$; the plain search takes about three minutes per case on
$128$ threads. The two lists coincide in all $18$ cases. Second, for $n=18$ and $a=1$ we
recorded where Theorem~\ref{thm:dual} fails for each of the three candidate projections
(Section~\ref{sec:family}). This takes about one second per projection.

\subsection{Switching closure (E5)}\label{sec:exp-switching}

Table~\ref{tab:switching} summarizes the switching computations of
Section~\ref{sec:switching}. In each dimension we computed the switching closure of the sources over switching paths whose
intermediate functions remain quadratic, as described in Section~\ref{sec:switching}. Every
nonzero class of every quotient $K(F,u)/T(F,u)$ was formed. Every quadratic neighbour that is
not a new class was certified EA-equivalent to a stored class. In dimensions eight, nine and
ten no quadratic class in the computed closure has a switching neighbour of degree three or
more. Among the power
functions only the Gold functions have nontrivial switchings, $x^3$ in all three dimensions and
$x^9$ in dimension eight. The Kasami, Welch, Niho, Dobbertin and inverse functions have none in
these dimensions. In dimension eight the closure has $17$ classes, none of which is
CCZ-equivalent to a permutation. This is consistent with \cite{DBLP:journals/tit/BeierleL22},
where switching was applied to all quadratic APN functions known in dimension eight. In
dimension ten the closure of the $25$ classes contains no other class. Only $x^3$, the two
F4 classes and four of the five classes of \cite{DBLP:journals/ccds/BudaghyanIK23} have
nontrivial switching neighbours. The classes of F3, F12, F13 and of
\cite{DBLP:journals/tit/BeierleL22} have none.

\begin{table}[t]
\caption{Exact switching closure (E5). ``Classes'' is the number of pairwise EA-inequivalent
quadratic classes in the closure; ``Neighb.'' is the number of functions formed, one per
nonzero class of a quotient $K(F,u)/T(F,u)$; ``Certified'' is the number of explicit
EA-equivalences proved, between a neighbour or a source and a stored class; ``deg $\ge3$'' indicates whether a class of the closure has a switching
neighbour of degree at least three (for non-quadratic sources such as the Kasami function, the
quotient is trivial in all our cases); ``Perm.'' gives the classes that are CCZ-equivalent to a
permutation.}\label{tab:switching}
\begin{tabular*}{\textwidth}{@{\extracolsep\fill}clrrrcl}
\toprule
$n$ & Sources & Classes & Neighb. & Certified & deg $\ge 3$ & Perm. \\
\midrule
6 & $x^3$, F4, Edel--Pott & 13 & 1{,}259 & 1{,}186 & yes (5) & 1 (Kim) \\
8 & powers, F4 & 17 & 96 & 83 & no & none \\
9 & powers, F4 & 6 & 14 & 12 & no & (odd $n$) \\
10 & powers, 25 classes & 25 & 13 & 18 & no & none \\
\botrule
\end{tabular*}
\end{table}

\subsection{Searching for new quadratic APN functions in dimension ten (E6--E8)}\label{sec:exp-new}

Since all $25$ classes examined here are excluded, an APN permutation CCZ-equivalent to a
quadratic function would have to come from a class outside this tested set. We tried three
generators. The first (E6) is a feasibility test in dimensions six to eight. It shows that
unstructured search does not scale, so we did not run it in dimension ten. The other two (E7,
E8) run in dimension ten.

\paragraph{Search over bilinear forms (E6).}
A quadratic function is APN if and only if its alternating bilinear form
$B(x,y)=F(x+y)+F(x)+F(y)+F(0)$ satisfies $B(x,y)\neq 0$ for all linearly independent $x,y$. We
build $B$ on $\langle e_1,\dots,e_k\rangle$ for increasing $k$. The extension by $e_{k+1}$
amounts to choosing a linear map $L(x)=B(x,e_{k+1})$ with $L(x)\notin B(x,\langle
e_1,\dots,e_k\rangle)$ for all nonzero $x$. We explore these choices by a randomized
depth-first search. In dimension six the search reaches all thirteen classes from $3{,}000$
restarts, in three seconds; their $13$ invariants are pairwise distinct, so every class is
reached. In dimension seven it finds $237$ solutions with $194$ distinct invariants, hence at
least $194$ classes, where $488$ classes are known \cite{DBLP:journals/ccds/KalginI23}. In dimension
eight it finds no solution in $36$ runs of $3\cdot10^7$ nodes. Unstructured search is
therefore not a viable generator beyond dimension seven. This is consistent with the use of
self-equivalences in \cite{DBLP:journals/tit/BeierleL22}.

\paragraph{Symmetry-restricted univariate families (E7).}
We sampled quadratic polynomials $\sum_e c_ex^e$, $e=2^i+2^j$, in $522$ families:
\begin{itemize}
\item coefficients in $\Ftwo$ (Frobenius-invariant functions), with four densities;
\item coefficients in the subfields $\F_4$ and $\F_{32}$;
\item sparse polynomials with two to five terms;
\item polynomials whose exponents lie in a single residue class modulo $m$, for each divisor
$m\le 341$ of $1023$ (functions with $F(\zeta x)=\zeta^dF(x)$ for $\zeta$ of order $m$).
\end{itemize}
For quadratic functions the APN property is decided from the $n\times n$ bilinear matrix in
about $40\,\mu$s, and faster for the many candidates rejected early. Of the $522$ families, $372$ residue families contain no exponent $2^i+2^j$, so they have
no candidates; the other $150$ families sample. Each of them got a maximum budget of $2{,}560$
batches of $2{,}000$ candidates, with a fixed seed. Before each batch a family stops if it has
already found $2{,}000$ APN functions, counted with repetitions; so it ends with fewer
than $4{,}000$, and all of them are kept. $56$ families stopped in this way, and $94$ used their
full budget. In total we tested $535{,}492{,}000$ candidates, out of a maximum of
$150\cdot 2{,}560\cdot 2{,}000=768{,}000{,}000$. They gave $166{,}157$ APN functions counted
with repetitions, and $105{,}595$ distinct ones; for instance, a family with a single exponent
contains at most $1{,}023$ distinct functions. The repository records these numbers for every
family, and lists every distinct function with its test result. Every distinct
APN function found was tested individually with Algorithm~\ref{alg:dual}. None is CCZ-equivalent to a permutation. Their
invariants are those of $11$ of the twenty classes of E4, and no other invariant occurs. The
run took $15$ minutes on $64$ server threads.

\paragraph{Tree search with prescribed self-equivalences (E8).}
We ran the deterministic tree search of Beierle and Leander \cite{Beierle20code} for each of
the $195$ classes of linear self-equivalences of $\F_2^{10}$ considered in
\cite{DBLP:journals/tit/BeierleL22}, with a time limit of five minutes per class. The search
terminated for $69$ classes. Solutions were found in six classes, giving $686$ APN functions with
nine distinct invariants, all among those of the twenty classes of E4. The test applied to each of the $686$
functions finds no CCZ-equivalent permutation. For the $126$ classes in which the search did
not terminate, we then ran the randomized version of the tree search for $24$ hours on $128$
threads. That is about one CPU-day per class ($6{,}144$ runs of $30$ minutes). Solutions were
again found only in the six classes above. Together with the deterministic runs we obtained
$4{,}638$ distinct APN functions with $16$ distinct invariants, all among those of the twenty
classes of E4. None of
the $4{,}638$ functions is CCZ-equivalent to a permutation. In particular, no solution was found
in the two classes for which \cite{DBLP:journals/tit/BeierleL22} reports that solutions must
exist but were not found after 16 CPU-days each.

\section{Conclusions and future work}\label{sec:conclusion}

We gave a parametrization of the Walsh zero spaces of functions with affine Walsh supports.
It turns the test for CCZ-equivalence to a permutation into searches for subspaces of $\Fn$
and for bilinear maps between them. This projection test decides the cases in which the
necessary condition of \cite{DBLP:journals/ffa/GologluL21,DBLP:journals/ccds/GologluP21} is
inconclusive. For listing all Walsh zero spaces it is $39$ to $55$ times faster than the
generic search of sboxU in dimension twelve, on the functions we measured. With it we settled
the family $x^3+a^{-1}\Tr(a^3x^9)$ in all even dimensions up to $24$, by computation for
$n\in\{6,10,14,18,22\}$. This includes a case in dimension $18$ that the known necessary
condition leaves open.

In dimension ten we checked the $25$ classes of quadratic APN functions represented by the
published instances we examined. Five of them, from
\cite{DBLP:journals/ccds/BudaghyanIK23}, are not in the earlier lists; for these our test
confirms the result obtained there with sboxU. We computed the quadratic switching closure of the
source functions listed in Table~\ref{tab:switching} in dimensions eight to ten, and tested every function found by two generators
of quadratic APN functions in dimension ten. No APN permutation was found. For $n=8$, the
inverse function fails the APN condition exactly on the two-dimensional subspaces $a\F_4$, and
swapping two entries of its lookup table never brings it closer to the APN property. The
butterflies with power-based keyed permutations, the cyclotomic maps and the switched inverse
functions that we enumerated contain no permutation of $\F_2^8$ closer to the APN property than
the inverse function.

Several directions remain open.
\begin{itemize}
\item \emph{Conjecture~\ref{conj:Fa}.} It holds in all computed cases, up to $n=22$. A proof
would settle the family $x^3+a^{-1}\Tr(a^3x^9)$ in all even dimensions. It needs two
ingredients. The first is a description of the subspaces of dimension $m-1$ and $m$ of the
cubes for odd $m$. The second is an analysis of the case $\dim(W\cap H_a)=m-1$, where the
argument of \cite{DBLP:journals/ccds/GologluP21} no longer applies.
\item \emph{Non-affine Walsh supports.} Theorem~\ref{thm:dual} requires affine supports.
Plateaued functions need not have affine supports. Proposition~\ref{cor:counting} still applies
to plateaued and general functions, so an extension of the projection test to them would be
useful. Natural test cases are the Kasami
function in dimension ten, which is not covered by \cite{DBLP:journals/ffa/GologluL21} since
$10\not\equiv0\pmod 4$, and the Dobbertin function $x^{339}$.
\item \emph{New quadratic APN functions in dimension ten.} The bottleneck of the CCZ approach
in dimension ten is the small number of known quadratic APN classes. The complete
classification of the functions with coefficients in $\F_2$ \cite{YuLIK25} gives no new class.
New classes therefore have to come from larger coefficient fields or other structures. On the functions we tested, the
test costs a fraction of a second per function. So any generator that produces many new
classes can be combined with it directly, for instance the constructions of Beierle et al.\
\cite{DBLP:journals/corr/abs-2508-04644} adapted to dimension ten.
\item \emph{Dimension twelve.} The two functions we measured in dimension twelve are decided
in under a second, by the dimension bound, and listing all their Walsh zero spaces takes about
a minute (Section~\ref{sec:exp-ccz}). Testing all known quadratic APN classes in dimension
twelve, and their switching closure, seems a plausible next step. The functions of Göloğlu
and Kölsch \cite{GologluK26} in dimension twelve would be natural test cases.
\item \emph{Two-step switching.} The switching graph of quadratic APN functions in dimension
ten is small. Switching simultaneously in two directions, $F+u_1f_1+u_2f_2$, leads to
quadratic rather than linear systems and is a natural next step.
\item \emph{Program search.} Romera-Paredes et al.\ recently used large language models to search for
programs that construct combinatorial objects
\cite{DBLP:journals/nature/RomeraParedesBNBKDREWFKF24}. A candidate construction of a
permutation, given as a program that works in every dimension, can be scored exactly by
$\cost$. The big APN problem is therefore a natural target for such methods.
\end{itemize}

\backmatter

\section*{Statements and Declarations}

\begin{itemize}
\item \textbf{Funding:} No dedicated funding was received for conducting this study.
\item \textbf{Competing interests:} The author has no competing interests to declare that
are relevant to the content of this article.
\item \textbf{Data and code availability:} All scripts, the C and Python sources, the
generated data and the expected results are publicly available at
\url{https://github.com/peacker/ccz-projection-test-apn} (release v1.0, commit \texttt{0ac04cd}). The repository contains one script per experiment, which recomputes its
results, and a script that compares every number reported in this paper with the stored
results.
\item \textbf{Use of AI tools:} The code, the experiments and a first draft
of the manuscript were prepared with the assistance of a large language model (Claude,
Anthropic). Section~\ref{sec:experiments} lists the checks applied to the results and what
each of them validates. The author takes full responsibility for the content.
\end{itemize}

\bibliography{references}

\end{document}